\documentclass[fleqn,usenatbib]{mnras}
\usepackage{newtxtext,newtxmath}
\usepackage{verbatim}

\usepackage[T1]{fontenc}

\DeclareRobustCommand{\VAN}[3]{#2}
\let\VANthebibliography\thebibliography
\def\thebibliography{\DeclareRobustCommand{\VAN}[3]{##3}\VANthebibliography}

\usepackage{graphicx}	% Including figure files
\usepackage{amsmath}	% Advanced maths commands
\usepackage{subfiles}   % Seperates files into subfiles for organisation
\usepackage{booktabs}   % what does this do?
\usepackage{natbib}     % Additional citation controls
\usepackage{orcidlink}
\usepackage{tabularx}
\usepackage{multirow}

\newcommand{\hii}{\mbox{H\,{\sc ii}}}

\newcommand{\hbeta}{\mbox{H\,{\sc $\beta$}}}
\newcommand{\halpha}{\mbox{H\,{\sc $\alpha$}}}
\newcommand{\hgamma}{\mbox{H\,{\sc $\gamma$}}}
\newcommand{\hdelta}{\mbox{H\,{\sc $\delta$}}}

\newcommand{\oiiia}{\mbox{[O\,{\sc iii]}{$\lambda 5007$}}}
\newcommand{\oiiib}{\mbox{[O\,{\sc iii]}{$\lambda 4363$}}}
\newcommand{\ariii}{\mbox{[Ar\,{\sc iii]}{$\lambda 7135$}}}

\newcommand{\arivdoublet}{\mbox{[Ar\,{\sc iv]}{$\lambda\lambda 4713,4741$}}}
\newcommand{\oii}{\mbox{[O\,{\sc ii]}{$\lambda\lambda 3726,3729$}}}
\newcommand{\sii}{\mbox{[S\,{\sc ii]}{$\lambda\lambda 6717,6731$}}}
\newcommand{\siia}{\mbox{[S\,{\sc ii]}{$\lambda 6717$}}}
\newcommand{\siib}{\mbox{[S\,{\sc ii]}{$\lambda 6731$}}}
\newcommand{\siiia}{\mbox{[S\,{\sc iii]}{$\lambda 6312$}}}

\newcommand{\auroraloii}{\mbox{[O\,{\sc ii]}{$\lambda\lambda 7322,7332$}}}

\newcommand{\neiii}{\mbox{[Ne\,{\sc iii]}{$\lambda 3870$}}}

\newcommand{\oiiinwl}{\mbox{[O\,{\sc iii]}}}
\newcommand{\oiinwl}{\mbox{[O\,{\sc ii]}}}
\newcommand{\feiinwl}{\mbox{[Fe\,{\sc ii]}}}

\newcommand{\arivnwl}{\mbox{[Ar\,{\sc iv]}}}

\newcommand{\siiinwl}{\mbox{[S\,{\sc iii]}}}

\title[The Ar/O and Ne/O ratios of $z\simeq4$ galaxies]{The JWST EXCELS survey: tracing the chemical enrichment pathways of high-redshift star-forming galaxies with O, Ar and Ne abundances}
\author[T. M. Stanton et al.]{
T. M. Stanton \orcidlink{0000-0002-0827-9769},$^{1}$\thanks{E-mail: t.stanton@ed.ac.uk}
F. Cullen \orcidlink{0000-0002-3736-476X},$^{1}$
A. C. Carnall,$^{1}$ 
D. Scholte \orcidlink{0000-0002-6867-1244},$^{1}$ 
K. Z. Arellano-C\'ordova \orcidlink{0000-0002-2644-3518},$^{1}$ \and \ 
D. J. McLeod \orcidlink{0000-0003-4368-3326},$^{1}$
R. Begley \orcidlink{0000-0003-0629-8074},$^{1}$ 
C. T. Donnan \orcidlink{0000-0002-7622-0208},$^{1}$  
J. S. Dunlop,$^{1}$  
M. L. Hamadouche \orcidlink{0000-0001-6763-5551},$^{2}$ 
R. J. McLure,$^{1}$ \and \
A. E. Shapley,$^{3}$
C. Bondestam,$^{1}$ and
S. Stevenson$^{1}$  
\\
$^{1}$Institute for Astronomy, University of Edinburgh, Royal Observatory, Edinburgh, EH9 3HJ, UK\\
$^{2}$Department of Astronomy, University of Massachusetts, Amherst, MA 01003, USA;\\
$^{3}$Department of Physics \& Astronomy, University of California, 430 Portola Plaza, Los Angeles CA 90095, USA\\
}

\date{Accepted XXX. Received YYY; in original form ZZZ}

\pubyear{2025}

\begin{document}
\label{firstpage}
\pagerange{\pageref{firstpage}--\pageref{lastpage}}
\maketitle

\begin{abstract} 
We present an analysis of eight star-forming galaxies with $\langle z \rangle = 4.0$ from the JWST EXCELS survey for which we obtain robust chemical abundance estimates for the $\alpha$-elements O, Ne and Ar.
The $\alpha$-elements are primarily produced via core-collapse supernovae (CCSNe) which should result in $\alpha$-element abundance ratios that do not vary significantly across cosmic time.
However, Type Ia supernovae (SNe Ia) models predict an excess production of Ar relative to O and Ne.
The ${\rm Ar/O}$ abundance ratio can therefore be used as a tracer of the relative enrichment of CCSNe and SNe Ia in galaxies.
Our sample significantly increases the number of sources with measurements of ${\rm O/Ar}$ at $z > 2$, and we find that our sample exhibits sub-solar Ar/O ratios on average, with $\rm{Ar/O} = 0.65 \pm 0.10 \, (\rm{Ar/O})_{\odot}$.
In contrast, the average Ne/O abundance is fully consistent with the solar ratio, with $\rm{Ne/O} = 1.07 \pm 0.12 \, (\rm{Ne/O})_{\odot}$.
Our results support a scenario in which Ar has not had time to build up in the interstellar medium of young high-redshift galaxies, which are dominated by CCSNe enrichment.
We show that these abundance estimates are in good agreement with recent Milky Way chemical evolution models, and with Ar/O trends observed for planetary nebulae in the Andromeda galaxy.
These results highlight the potential for using multiple element abundance ratios to constrain the chemical enrichment pathways of early galaxies with JWST.
\end{abstract}


%%%%%%%%%%%%%%%%%%%%%%%%%%%%%%%%%%%%%%%%%%%%%%%%%%

%%%%%%%%%%%%%%%%% BODY OF PAPER %%%%%%%%%%%%%%%%%%

\section{Introduction}
\label{sec:intro}

    By tracing the abundances of heavy elements within the interstellar medium (ISM), we can constrain the key secular processes regulating the growth of galaxies, namely the history of star formation and the impact of large-scale inflows and outflows of gas \citep{Maiolino_Mannucci_2018, Peroux_2020}.
    Measurements of O, the most abundant metal by mass, provide a good proxy for the total metallicity and are readily measurable from bright rest-frame optical lines.
    These have been used to establish the existence of the gas-phase mass-metallicity relationship (MZR; \citealp{Tremonti_2004, erb2006, Maiolino_2008, Andrews_2013, cullen_2014, Sanders_2021, scholte2024}) and investigate the fundamental metallicity relationship (FMR; \citealp{LaraLópez+2010, Mannucci+2010, Sanders_2021, Curti+2023}) from the local Universe out to $z \simeq 10$.
    
    Further insights can be gained by tracing the relative abundances of different elements \citep[e.g.,][]{Kobayashi_2020}.
    For example, studies at $z > 2$ have increasingly focused on measurements of the stellar Fe abundance from ultra-deep rest-frame far-ultraviolet (FUV) data. 
    These analyses have revealed the existence of the stellar MZR at high redshifts \citep[e.g.,][]{Cullen_2019, Kashino_2022, Chartab_2023, Stanton_2024}.
    It has also become clear that star-forming galaxies at these redshifts exhibit elevated O/Fe ratios relative to the solar values (i.e., $\alpha-$enhancement), a sign that the interstellar medium (ISM) is enriched predominantly by the products of core-collapse supernovae (CCSNe).
    This is a direct consequence of the delayed production of iron via Type-Ia supernovae (SNe Ia) - with a typical delay time of $\simeq 1 \, \rm{Gyr}$ - resulting in an iron-deficient ISM \citep{Kobayashi_2020}.
    Most recently, in \citet{Stanton_2024} we simultaneously measured the O and Fe abundances of $65$ star-forming galaxies at $\langle z \rangle \simeq3.5$, finding an average enhancement of ${\langle \mathrm{O/Fe} \rangle \simeq 2.7 \times \mathrm{(O/Fe)_\odot}}$, consistent with other measurements across the literature \citep[e.g.,][]{Steidel_2016, Topping_2020A, Topping_2020B, Cullen_2021, Kashino_2022, Strom_2022, Chartab_2023}.
    
    These $\alpha$-enhanced abundances offer a natural explanation for the offset observed between high-redshift and local populations in strong emission-line ratio diagrams \citep[e.g., the BPT diagram;][]{Shapley_2019, Jeong_2020, Clarke_2023, Shapley_2024}.  
    For example, an Fe deficit will result in harder stellar ionising spectra at fixed O abundance \citep[e.g.,][]{Topping_2020A}, thereby increasing high-ionisation line strengths (e.g. an elevated \oiiinwl/\hbeta \ ratio).
    Furthermore, in \citet{Stanton_2024} we have demonstrated how interpreting the mass-dependence of element abundance ratios such as ${\rm O/Fe}$ with chemical evolution models \citep[e.g.,][]{Weinberg+2017} can place unique constraints on galaxy-scale outflow efficiencies \citep[see also;][]{Chartab_2023}.
    Perhaps most intriguingly, abundance ratios such as O/Fe also represent a unique pathway for linking high-redshift galaxies with detailed stellar archaeological measurements of the Milky Way and Local Group (e.g., \citealp{Cullen_2021}, Monty et al. in prep)
    connecting star formation at high-redshift to descendant populations within our own Galaxy.

    Despite their growing number, estimates of Fe abundances at high redshift remain extremely difficult to obtain because of the requirement for rest-frame FUV continuum spectroscopy.
    Although progress may be made for the brightest sources at extreme redshifts with JWST \citep[e.g.,][]{nakane2024}, individual Fe determinations will not be possible for large samples of galaxies until the Extremely Large Telescope era.
    Whilst \feiinwl \ lines are measurable in the rest-frame optical and near-IR, interpreting measurements of these features requires significant dust depletion corrections. 
    However, in lieu of using Fe, other elements with SNe Ia production channels can be used to trace the relative enrichment of SNe Ia  and CCSNe at early times.
    One such promising element is Ar.
    Although Ar is an $\alpha$-element, models predict that $\simeq 34$ per cent of the total Ar production is a result of SNe Ia \citep{kobayashi_2020a}.
    As a result, the O/Ar ratio can be used as a tracer of the degree of enrichment of CCSNe  versus SNe Ia, and, following Fe, one would expect a deficit of Ar with respect to O in high-redshift galaxies.
    
    Indeed, observational evidence for delayed Ar production was recently provided by \citet{Arnaboldi+2022}, who analysed the abundances of O and Ar in planetary nebulae and \hii \ regions within M31 and found an increase in ${\rm O/Ar}$ towards lower ${\rm Ar/H}$.
    The observed sequence in ${\rm O/Ar}-{\rm Ar/H}$ displays a striking similarity to the ${\rm O/Fe}-{\rm Fe/H}$ sequence observed in Milky Way stars.
    Moreover, a small number of recent studies have demonstrated that it is possible to constrain direct Ar abundances of individual galaxies at high redshift from detections of the \ariii \ and \arivdoublet \ emission lines.
    For example, \citet{Rogers_2024} performed a thorough chemical abundance analysis of a $z\sim3$ galaxy observed as part of the CECILIA survey \citep{strom_2023}, finding a sub-solar ${\rm Ar/O}$ abundance ratio, consistent with a scenario wherein Ar has a significant SNe Ia production channel. 
    In contrast, an analysis by \citet{Welch_2024} of the Sunburst Arc (a highly lensed galaxy at $z\simeq2.37$) yielded an ${\rm O/Ar}$ measurement close to the solar value.
    Recently, \citet{Bhattacharya+2024} have analysed a sample of public-release JWST spectra at $1.3 < z < 7.7$, finding ${\rm O/Ar}$ ratios that are mostly consistent with a combination of CCSNe and SNe Ia production channels. 

    These early JWST studies clearly demonstrate the potential of using Ar to study the chemical enrichment and star-formation histories of galaxies in the early Universe out to $z \simeq 6$ (beyond which most Ar lines emission features are inaccessible with JWST/NIRSpec).
    In contrast to Fe, which requires deep stellar continuum observations, Ar abundances can be estimated using emission lines that are readily detected by JWST.
    However, the sample of high-redshift galaxies with direct Ar abundance determinations remains small, and the current results do not paint an entirely clear picture. 
    A full understanding the formation channels of Ar, and its potential as a tracer of $\alpha$-enhanced abundances at early times, requires larger samples.
    
    In this work, we select a sample of star-forming galaxies from the JWST Early eXtragalactic Continuum and Emission Line Survey \citep[EXCELS;][]{Carnall_2024} in the redshift range $1.8 < z < 5.3$, with detections of the emission lines required to directly measure the abundances of O and Ar.
    Crucially, we are also able to constrain the Ne abundances for our sample using the \neiii \  emission line.
    As a pure $\alpha$-element, the enrichment of Ne is expected to follow O and therefore can be used as a control element.
    Specifically, we can test the hypothesis that while Ne/O ratios should be consistent with the solar value in high redshift galaxies, Ar/O ratios might show an offset to lower values.
    
    The structure of this paper is as follows. 
    In Section~\ref{sec:data}, we discuss our sample selection and the EXCELS data set used in this work.
    The methodologies we employ to measure the abundances of O, Ar and Ne from our spectra are described in Section~\ref{sec:analysis}.
    We present our measured abundance ratios and discuss our results in Section~\ref{sec:results}. 
    In Section~\ref{sec:discussion}, we discuss the effect of systematic uncertainties related to atomic data, ionisation correction factors (ICFs), and dust depletion on our results.
    Finally, we summarise our conclusions in Section~\ref{sec:conclusions}.

    Throughout this work, metallicities are quoted relative to a solar abundance value taken from \citet{Asplund_2021} unless otherwise stated.
    The key solar abundance values relevant to our analysis are: $12 + \log(\mathrm{O/H}) = 8.69$, $12 + \log(\mathrm{Ne/H}) = 8.06$ and $12 + \log(\mathrm{Ar/H}) = 6.38$. 
    We assume the following cosmology: $\Omega_{\rm M} = 0.3$, $\Omega_{\rm \Lambda} = 0.7$, and $H_0 = 70 \, {\rm km \, s^{-1} \, Mpc^{-1}}$.

\section{Data and Sample Properties}
\label{sec:data}

        The primary galaxy sample analysed here is taken from the JWST EXCELS spectroscopic survey (GO 3543; PIs: Carnall, Cullen; \citealp{Carnall_2024}).
        The EXCELS survey is designed to provide medium-resolution ($\mathrm{R}=1000$) spectroscopy of targets selected from JWST/PRIMER imaging (GO 1837; PI: Dunlop) in the UDS field. 
        We observe 4 NIRSpec/MSA pointings that are optimised to target rare massive quiescent galaxies at high redshift, with the remaining MSA slitlets allocated to star-forming galaxies in the redshift range $1 < z < 8$.
        Full details of the EXCELS target selection, observing strategy, and data reduction are described in \citet{Carnall_2024}.
        Below, we briefly review the observations and data reduction.

    \subsection{Data reduction}
    
        Each of the four NIRSpec/MSA pointings is observed with the G140M/F100LP, G235M/F170LP and G395M/F290LP gratings.
        Observations were conducted using a 3 shutter slitlet and 3-point dither pattern.
        The total integration times within each grating for each pointing are $\simeq 4$ hours in G140M and G395M, and $\simeq 5.5$ hours in G235M, using the NRSIRS2 readout pattern.
        As described in \citet{Carnall_2024}, separate MSA configurations are specified for each grating to maximise the number of objects for which we can observe the key rest-frame UV and optical spectral features. 
        The majority of objects presented in this work have observations in the G235M and G395M gratings which provides coverage of the vital rest-frame optical emission lines at the median redshift of our sample.

        To reduce the data, we begin by processing the raw level 1 products from the Mikulski Archive for Space Telescopes (MAST) using v1.15.1 of the JWST reduction pipeline\footnote{https://github.com/spacetelescope/jwst}.
        We run the default level 1 configuration except for turning on advanced snowball rejection and make use of of the CRDS$\_$CTX = jwst$\_$1258.pmap version of the JWST Calibration Reference Data System (CRDS) files.
        We then run the level 2 pipeline steps assuming the default configuration except for turning off the sky subtraction step.
        The level 3 pipeline provides the final combined 2D spectra that can be used for science analysis.
        We then perform our own custom 1D optimal extraction of the 2D spectra \citep{Horne_1986} setting the extraction centroid as the flux-weighted mean position of the object within the NIRSpec/MSA slitlet \citep[see][]{Carnall_2024}.

        A flux-calibration step is needed to correct for wavelength-dependent slit losses due to the increasing FWHM of the PSF with wavelength.
        Additionally, small offsets in the source position within the slit can result in flux mismatches between the different gratings.
        To match the flux level between the different gratings, we calculate the median flux in overlapping wavelength regions and scale the G140M and/or G395M spectra to the G235M spectra.
        
        To achieve the absolute flux calibration we perform a spectrophotometric calibration to the broadband photometry similar to the approach adopted for the CECILIA survey \citep{Rogers_2024}.
        Each galaxy in the EXCELS sample benefits from deep HST and JWST/NIRCam imaging spanning the wavelength range $0.4 - 5 \mu \mathrm{m}$ in 11 photometric bands: F435W, F606W, F814W, F090W, F115W, F150W, F200W, F277W, F356W, F410M and F444W.
        Broadband fluxes from the PSF-homogenised images are extracted using $0.5$ arcsec diameter apertures and corrected to total using the F356W FLUX$\_$AUTO values measured by \textsc{SourceExtractor} \citep{bertin_sextractor}.
        We integrate each EXCELS spectrum through the overlapping filters and scale these to the imaging fluxes.
        The correction as a function of wavelength is derived via a linear interpolation between the bands.
        As all of the sources studied in this paper have significant stellar continuum detections (e.g., Fig \ref{fig:example-galaxy}) scaling to the photometry in this way should account for wavelength-dependent slit losses and provide as accurate a flux calibration as is possible.
        Applying this method to the four quiescent galaxies presented in \citet{Carnall_2024} we find that the flux calibration is accurate to within $\lesssim 5$ per cent\footnote{These quiescent galaxies benefit from high S/N continuum detections and well-understood continuum properties. This enables more accurate flux calibrations to be performed via comparisons to empirical spectral templates as described in \citet{Carnall_2024}. Our cruder approach (i.e., a simple spectro-photometric scaling) yields fluxes accurate to within $\simeq 5$ per cent of this method.}.

    \subsection{Sample Selection}

        To achieve our science goals we require detections of all of the emission lines necessary for a direct determination of O, Ar and Ne abundances.
        From the full EXCELS sample of star-forming galaxies, we select galaxies with ${\rm S/N > 2}$ detections of:
        \begin{enumerate}
            \item The \oiiia, \oiiib \ emission lines necessary to directly measure the electron temperature ($T_e \, \mathrm{\textsc{[O iii]}}$) and the total oxygen abundance ($12 + \log{\rm (O/H)}$)
            \item The \ariii \ and \neiii \ emission lines required to determine ${\rm Ar/H}$ and ${\rm Ne/H}$
            \item At least two Balmer emission lines (i.e., \halpha, \hbeta, \hgamma, \hdelta) which we need to perform robust nebular reddening corrections.
        \end{enumerate}

        \begin{figure*}
            \centering
            \includegraphics[width=\linewidth]{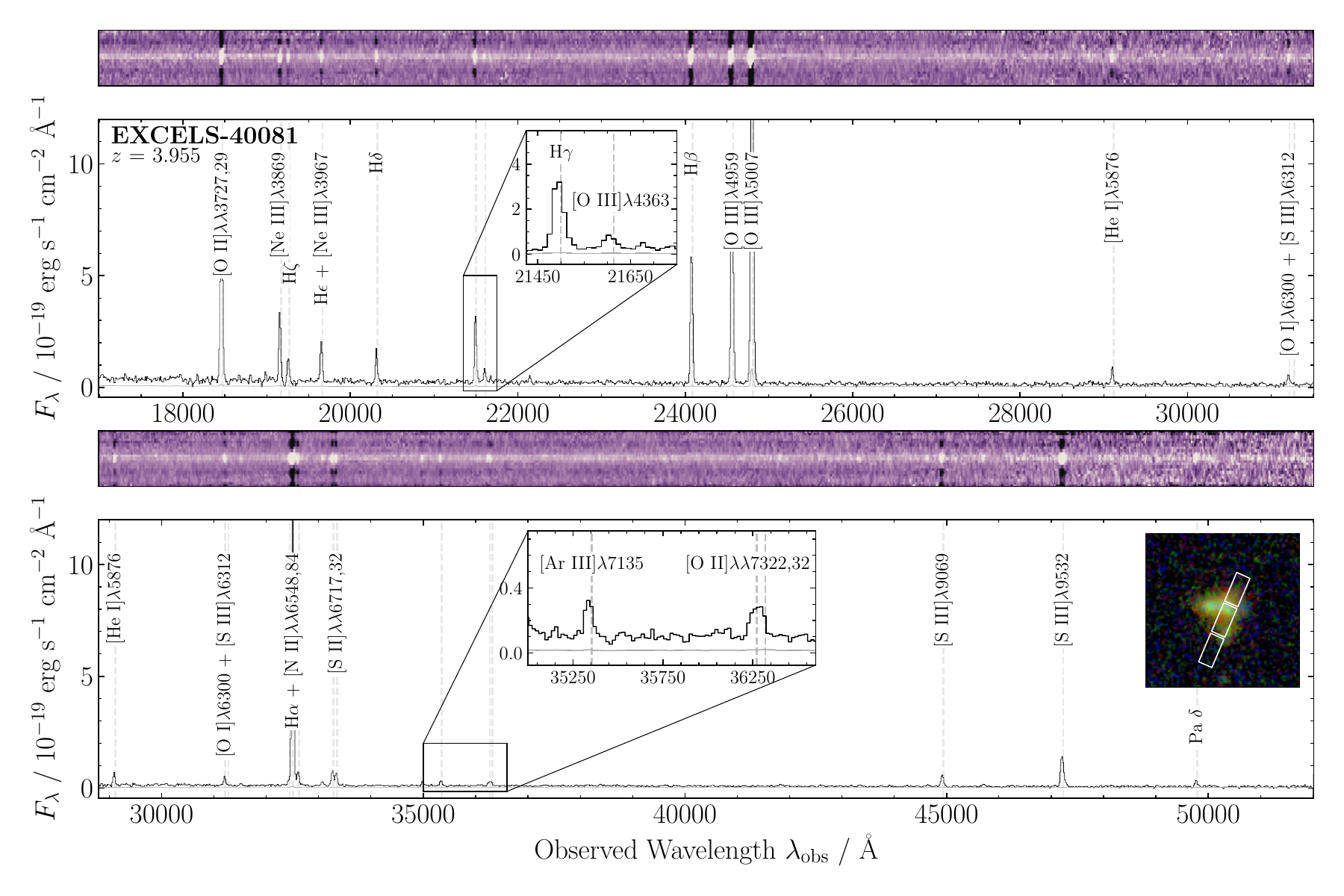}
            \caption{An example of one galaxy from our sample (EXCELS-40081, $z=3.955$) which features all of the required emission lines needed to determine the ${\rm Ar/O}$ and ${\rm Ne/O}$ abundance ratios. 
            The top two panels show the NIRSpec/G235M 2D and 1D spectra of this source, and the bottom two panels show the same for NIRSpec/G395M. 
            Prominent emission lines are labelled in each 1D spectrum. 
            Within the NIRSpec/G235M 1D panel, the inset panels show the \oiiib \ auroral emission line used to derive the \oiiinwl \ region electron temperature that is needed for robust element abundance estimates.
            Similarly in the NIRSpec/G235M 1D panel, the inset panel shows the \ariii \ line which we use to determine Ar abundances.
            In the upper right-hand corner of the bottom panel, we show a false three-colour image of EXCELS-40081 generated from the F115W+F150W, F200W+F277W and F356W+F444W PRIMER imaging with the NIRSpec/MSA slitlets overplotted in white.}
            \label{fig:example-galaxy}
        \end{figure*}
    
        From an initial sample of ten, we find two galaxies display \halpha/\hbeta \ ratios which are lower than the theoretical ratios from Case B recombination theory with a significance of $\gtrsim 2\sigma$.
        Deviations from Case B recombination ratios have been observed in a number of high-redshift galaxy spectra, and have been attributed to gas geometry and optical thickness to \halpha \ photons \citep[e.g.,][]{Scarlata_2024}, or alternatively to density bounded nebulae \citep[]{McClymont_2024}.
        Due to the increased uncertainty in modelling the ionization conditions and reddening corrections for these galaxies, we exclude them from our final sample.
        Our final sample therefore comprises of eight galaxies across a redshift range $1.8 < z < 5.3$.
        This sample represents a significant increase in the number of galaxies with direct ${\rm Ar/O}$ determinations at high redshift.
        An example galaxy from our final sample (EXCELS-40081) is shown in Figure~\ref{fig:example-galaxy}.

    \begin{table*}
\caption{ 
The IDs, redshifts, coordinates, masses and star-formation rates of the eight selected EXCELS star-forming galaxies.
The median redshift of the sample is $\langle z \rangle \simeq 4.0$. Stellar masses and SED-based star-formation rates are derived from fitting the emission-line corrected broad-band photometry with \textsc{Bagpipes} \citep{Carnall_2018}.
${\rm E(B-V)}$ values and \halpha-based SFRs are measured as described in Sections~\ref{subsec:dust-correction} and \ref{subsec:sfms}.}
\label{tab:ar-objects}
\renewcommand{\arraystretch}{1.3}
\setlength{\tabcolsep}{1pt}
\begin{tabularx}{\linewidth}{@{\extracolsep{\fill}}lccccccc@{}}
\toprule
EXCELS ID & RA & DEC & $z_{\mathrm{spec}}$ & $\log (M_\star / \mathrm{M}_\odot) $ & $\log (\mathrm{SFR_{SED}}/ \mathrm{M}_\odot \mathrm{yr}^{-1})$ & $\log (\mathrm{SFR_{H\alpha}} / \mathrm{M}_\odot  \mathrm{yr}^{-1})$ & ${\rm E(B-V)}$ \\ \midrule \midrule
EXCELS-40081 & $34.3659$ & $-5.2608$ & $3.954$ & $9.07^{\,+0.10}_{-0.09}$ & $1.52^{\,+0.14}_{-0.12}$ & $1.78\pm0.11$ & $0.05\pm0.07$ \\
EXCELS-45052 & $34.3671$ & $-5.2520$ & $4.234$ & $9.69^{\,+0.05}_{-0.05}$ & $2.08^{\,+0.12}_{-0.11}$ & $1.55\pm0.14$ & $0.32\pm0.10$ \\
EXCELS-45177 & $34.3632$ & $-5.2523$ & $2.901$ & $8.28^{\,+0.05}_{-0.06}$ & $1.21^{\,+0.08}_{-0.08}$ & $1.02\pm0.09$ & $0.12\pm0.03$ \\
EXCELS-47143 & $34.2622$ & $-5.2489$ & $3.233$ & $8.76^{\,+0.09}_{-0.12}$ & $0.88^{\,+0.26}_{-0.35}$ & $0.84\pm0.10$ & $0.14\pm0.03$ \\
EXCELS-52422 & $34.2503$ & $-5.2406$ & $4.025$ & $8.12^{\,+0.08}_{-0.08}$ & $0.73^{\,+0.11}_{-0.11}$ & $1.17\pm0.10$ & $0.07\pm0.06$ \\
%EXCELS-52712 & $34.3762$ & $-5.2401$ & $3.238$ & $8.71^{\,+0.29}_{-0.04}$ & $1.63^{\,+0.09}_{-0.07}$ & $1.35\pm0.09$ & $0.16\pm0.03$ \\
EXCELS-59720 & $34.3797$ & $-5.2282$ & $4.367$ & $9.31^{\,+0.10}_{-0.11}$ & $1.22^{\,+0.12}_{-0.10}$ & $1.15\pm0.09$ & $0.00\pm0.03$ \\
EXCELS-94335 & $34.2728$ & $-5.1717$ & $1.812$ & $8.95^{\,+0.07}_{-0.05}$ & $1.63^{\,+0.13}_{-0.19}$ & $1.19\pm0.09$ & $0.22\pm0.01$ \\
EXCELS-121806 & $34.4039$ & $-5.1304$ & $5.226$ & $8.11^{\,+0.09}_{-0.05}$ & $1.07^{\,+0.08}_{-0.07}$ & $1.25\pm0.09$ & $0.01\pm0.04$ \\ \bottomrule
\end{tabularx}
\end{table*}

    \subsection{Measurements and derived quantities}

        \subsubsection{Fitting emission-line profiles} \label{sec:fitting-em-lines} 

        To fit our emission lines we adopt a similar approach as used for the DESI EDR catalogue, \citep[\textsc{FastSpecFit},][]{Moustakas+2023}, a full description of which will be outlined in Scholte et al. (in prep).
        The emission line fluxes are measured using a two step approach where first the continuum flux is subtracted and following that all emission lines are fitted using Gaussian line profiles. 
        The continuum flux is measured using the running mean value of the 16$^{\rm th}$ to 84$^{\rm th}$ percentile flux values within a top hat function with a rest-wavelength width of 350 \AA. 
        This procedure masks out any strong emission features and returns the underlying continuum.
        We use the continuum-subtracted residuals to re-estimate the flux uncertainties returned by the analysis pipeline. 
        We do this by a multiplication factor such that $\tilde{\sigma}_{F} = \frac{1}{2}(R_{84} - R_{16})$, where $\tilde{\sigma}_{F}$ is the median flux uncertainty and $R_{16}$, $R_{84}$ are the 16$^{\rm th}$, 84$^{\rm th}$ percentiles of the residuals. 
        We find error multiplication factors in the range $\tilde{\sigma}_{F} \simeq 1.5 - 2$, in good agreement with independent analyses \citep[e.g.,][]{maseda_2023, Carnall_2024}.
        We fit the emission lines to the continuum subtracted residuals with re-estimated uncertainties using Gaussian line profiles. 
        All of the emission lines are fit simultaneously using a common intrinsic line width and line velocity, and the line amplitude is freely fitted for each emission line. 
        The total width of an emission line, $\sigma_{\rm tot}$, is a convolution of intrinsic broadening as well as instrumental broadening which is dependent on the line wavelength and the NIRSpec grating used: $\sigma_{\rm tot}^2 = \sigma_{\rm intrinsic}^2 + \sigma_{\rm grating}^2(\lambda)$. 
        The $\chi^2$-minimisation is performed with the \textsc{SciPy} least squares implementation \citep{Branch+1999, Scipy+2020}. 
        To determine the uncertainty on a given emission-line flux, we propagated the individual pixel errors weighted by their contribution to the Gaussian line flux.

        \subsubsection{Measuring stellar masses and star-formation rates} \label{sec:sed-fitting}
    
        To determine the masses and star-formation rates (SFRs) of our galaxies, we employ spectral energy distribution (SED) fitting  to the broadband photometry of our galaxies. 
        Each galaxy has full PRIMER imaging coverage from $\lambda_{\rm obs} = 0.4 - 5.0 \, {\rm \mu m}$, with approximately half of the sources having additional detections in the MIRI/F770W filter.
         
        We fit the broadband photometry with \textsc{Bagpipes} \citep{Carnall_2018} to determine the best-fitting SEDs, and the corresponding stellar masses and SFRs.
        We perform emission-line corrections to the photometry and fit only the stellar SED component.
        We fit with the \citet{Bruzual_Charlot_2003} models, and assume a flexible dust attenuation law following \citet{Salim_2018} and a double power-law star-formation history with a \citet{kroupa2001} initial mass function (IMF).
        Full details of the \textsc{Bagpipes} fitting for the EXCELS star-forming sample will be given in Scholte et al. (in prep.).

        The best-fitting stellar masses and star-formation rates (SFRs) and their associated uncertainties are calculated by taking the $16^\mathrm{th}$, $50^\mathrm{th}$ and $84^\mathrm{th}$ percentiles of their respective resulting posterior distributions.
        For the SFRs, we adopt values averaged over the last $10$ Myr of the star-formation history.
        The coordinates, redshifts, stellar masses and SED-based SFRs of our final sample are summarised in Table~\ref{tab:ar-objects}.
        For each galaxy, we measure the underlying stellar absorption of the Balmer lines from the best-fitting SED and correct our measured emission line fluxes for this absorption.
        These corrections result in median increases of $\lesssim 1$ per cent to the observed emission-line fluxes for the main Balmer lines used this study (\halpha \ and \hbeta).

        \subsubsection{Reddening corrections for nebular line fluxes} \label{subsec:dust-correction}

        In each of our galaxies we detect \halpha \ and \hbeta, with additional detections of \hgamma \ and \hdelta \ in some galaxies, and use the ratios of these lines to correct our observed emission-line fluxes for the effects of nebular dust reddening\footnote{For some galaxies we also detect lower order Balmer lines (e.g., ${\rm H\epsilon}$ and ${\rm H\eta}$), however, as these lines are contaminated by nearby emission lines, we do not include them in our nebular reddening calculation.}.
        For each Balmer line ratio, we calculate the theoretically expected line value from \citet{Storey_Hummer_1995} assuming case B recombination with $T_e = 12,000 \, \mathrm{K}$ and $n_e = 300 \, \mathrm{cm^{-3}}$.
        These values are chosen as the approximate average values measured across our sample .
        To calculate ${\rm E(B-V)}_{\rm{neb}}$ we take the inverse-variance weighted average across all Balmer ratios, which we ensure is consistent with each individual Balmer ratio, assuming the \citet{Cardelli_1989} extinction curve.
        We note that although the \citet{Cardelli_1989} curve has been shown to be a good approximation of the nebular attenuation law in high-redshift star-forming galaxies \citep[e.g.][]{Reddy_2020}, a recent detailed analysis of a deep galaxy spectrum at $z=4.41$ has demonstrated that the true attenuation law is likely to vary on a galaxy-by-galaxy basis \citep{sanders_2024}.
        We lack the dynamic range in wavelength to accurately the derive the dust curve for each of our objects individually, and therefore adopt the \citet{Cardelli_1989} relation as representative of the typical attenuation curve at these redshifts.
        To estimate the uncertainties, we perturb the line ratios by their associated errors 500 times and recalculate ${\rm E(B-V)}_{\rm{neb}}$, taking the median and standard deviation of the resulting distributions.
        The final values and uncertainties are reported in Table~\ref{tab:ar-objects}.
        We use this final value of ${\rm E(B-V)}_{\rm{neb}}$ to dust-correct all of our observed line fluxes.

        \subsection{\texorpdfstring{H$\boldsymbol{\alpha}$}{Ha} star-formation rates and the relation to the star-forming main sequence} \label{subsec:sfms}

        \begin{figure}
            \centering
            \includegraphics[width=0.8\linewidth]{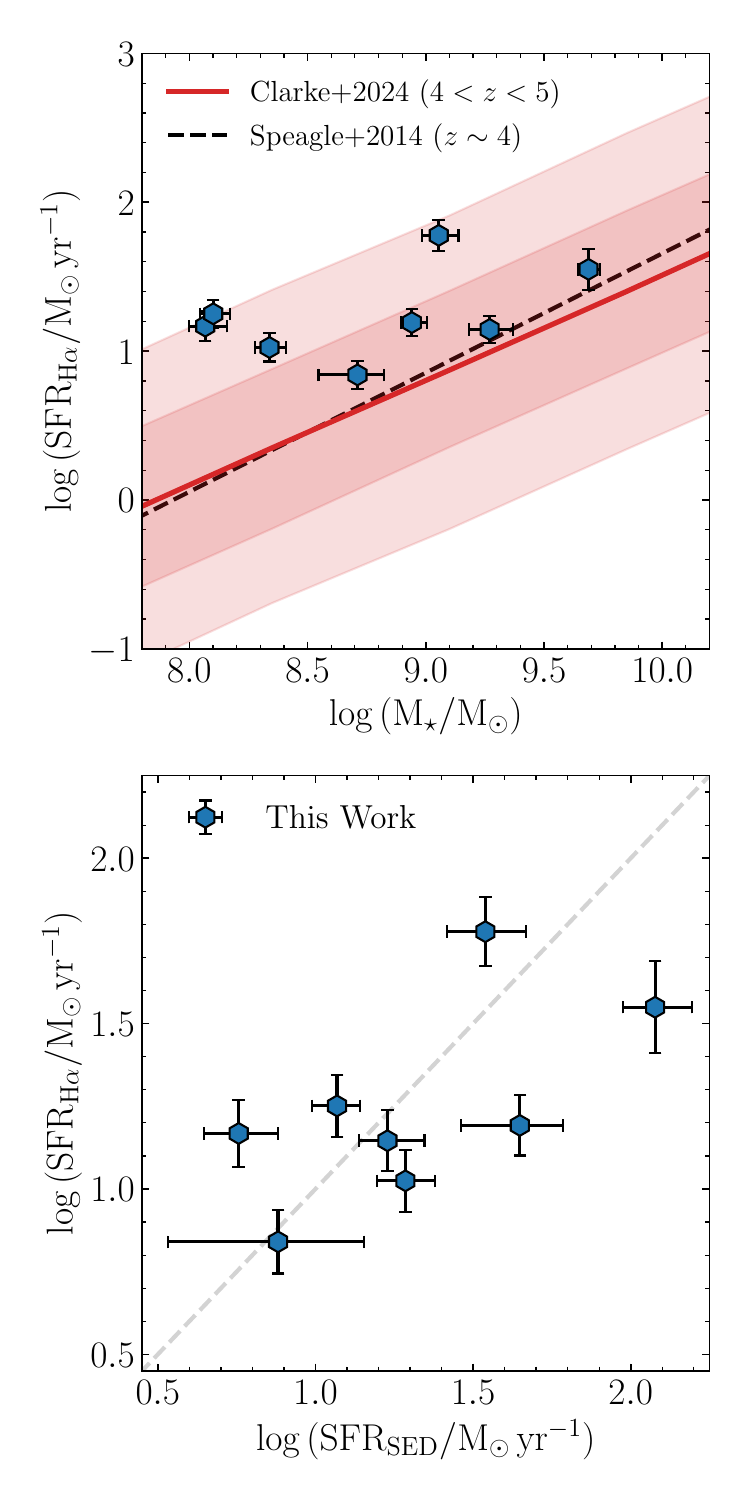}
            \caption{The top panel shows the star-forming main sequence for our sample (hexagons).
            The star-formation rates are derived from dust-corrected \halpha \ luminosities assuming the \citet{Hao_2011} conversion. 
            For comparison, we plot the star-forming main sequence relations from \citet{Speagle_2014} at $z=4$ (dashed black line) and from \citet{clarke2024} at $4 < z < 5$ (red line).
            The shaded red regions highlight the $1\sigma$ and $2\sigma$ intrinsic scatter intervals about the \citet{clarke2024} relation.
            The bottom panel shows a comparison between the \halpha- and SED-derived SFRs.
            The sample is generally scattered around the unity line (grey dashed) indicating that our SED fitting is accurately capturing the star-formation properties of the sample.
            Assuming either SFR estimate, it is clear that our sample consists of galaxies with above-average SFRs at $z \simeq 4$ (i.e., $\rm{SFRs} \, \gtrsim 10 \, \mathrm{M_{\odot}yr}^{-1}$).
            }
            \label{fig:star-forming-main-sequence}
        \end{figure}

        We convert our dust-corrected \halpha \ line luminosities into SFRs using the conversion of \citet{Hao_2011} for a \citet{kroupa2001} IMF \footnote{This conversion is chosen to maintain consistency with our SED modelling assumptions, though we note that other conversion factors tuned to sub-solar metallicities will return lower ${\rm SFR_{H\alpha}}$ \citep[e.g.,][]{clarke2024}, and as such our ${\rm SFR}$s may be overestimated.}.
        The bottom panel of Fig.~\ref{fig:star-forming-main-sequence} shows the comparison between the \halpha- and SED-derived SFRs, where it can be seen that there is good agreement.
        Taking into consideration the various systematic uncertainties and SED modelling assumptions (not included in Fig.~\ref{fig:star-forming-main-sequence}) we consider both estimates to be fully consistent.
        Assuming either SFR estimate, it is clear that the galaxies in our sample are highly star forming, with SFRs $\gtrsim 10 \, \mathrm{M_{\odot}yr}^{-1}$.
        
        In the upper panel of Fig.~\ref{fig:star-forming-main-sequence}, we show the positions of our sample with respect to star-forming main sequence at $z\simeq4-5$ (taken from \citealp{Speagle_2014} and \citealp{clarke2024}).
        It can be seen that the galaxies presented here sit primarily above the main sequence, although all fall approximately within the $2\sigma$ intrinsic scatter of the \citet{clarke2024} relation.
        This is the case regardless of whether we assume \halpha-  or SED-based SFRs.
        Therefore, we note that the results presented below are primarily applicable to the galaxies with above-average SFRs at $z\simeq4$, although it is worth pointing out that $4$ galaxies in our sample fall on, or within $1 \sigma$ of, the main sequence.

\section{Determining physical conditions and abundances} 
\label{sec:analysis}

        In order to measure the chemical abundances of our galaxies we first characterise their temperature and density structures.
        We assume a three-zone structure, comprising low-, intermediate-, and high-ionization zones \citep[e.g.,][]{Berg_2021}.
        We do not consider a fourth `very high' ionization zone because we do not significantly detect any ionic species corresponding to this zone.
        We attribute each ionic species to an ionization zone on the basis of its ionization potential (IP).
        The low ionization zone comprises $\mathrm{O^+}$ and $\mathrm{S^+}$, which have IPs of $13.6 \, \mathrm{eV}$ and $10.4 \, \mathrm{eV}$ respectively. 
        The ions $\mathrm{Ar^{2+}}$ and $\mathrm{S^{2+}}$, which have IPs of $27.6 \, \mathrm{eV}$ and $23.3 \, \mathrm{eV}$, are classified in the intermediate ionization zone.
        Finally, the high ionization zone contains $\mathrm{O^{2+}}$ and $\mathrm{Ne^{2+}}$, which have ionization potentials of $35.1 \, \mathrm{eV}$ and $41.0 \, \mathrm{eV}$.

        \begin{table}
\caption{The atomic data utilised in this work.}
\label{tab:pyneb-data}
\renewcommand{\arraystretch}{1.3}
\setlength{\tabcolsep}{1pt}
\begin{tabularx}{\linewidth}{@{\extracolsep{\fill}}cll@{}}
\toprule
Ion        & Transition Probabilities ($A_\mathrm{ij})$ & Collision Strengths ($\Upsilon_\mathrm{ij})$ \\ \midrule \midrule
O$^{+}$    & \citet*{Froese_Fischer_Tachiev_2004}        & \citet{Kisielius_2009} \\
O$^{2+}$   & \citet*{Froese_Fischer_Tachiev_2004}        & \citet*{Storey_2014} \\
Ne$^{2+}$   & \citet{Froese_Fischer_Tachiev_2004}        & \citet{McLaughlin_2011}  \\
S$^{+}$    & \citet*{Irimia_Froese_Fischer_2005}         & \citet*{Tayal_Zatsarinny_2010}   \\
Ar$^{2+}$  & \citet*{Mendoza_Zeippen_1983}               & \citet{Munoz_Burgos_2009} \\
\bottomrule
\end{tabularx}
\end{table}

        \subsection{Temperature structure} \label{subsec:temperature-structure}

        \begin{table*}
\caption{The temperatures, densities, and O/H, Ne/O and Ar/O abundance ratios for our sample. All $T_e \,$\siiinwl \ and $T_e \,$\oiinwl \ value are estimated from $T_e \,$\oiiinwl \ using equations~\ref{eq:garnett+1992} and \ref{eq:campbell+1986} respectively. The O abundances are uncorrected for dust depletion, as discussed in Section~\ref{sec:dust-depletion}.}
\label{tab:abundances}
\renewcommand{\arraystretch}{1.3}
\setlength{\tabcolsep}{1pt}
\begin{tabularx}{\linewidth}{@{\extracolsep{\fill}}lccccccr@{}}
\toprule
EXCELS ID & $T_e\,$[O III] / K & $T_e\,$[S III] / K & $T_e\,$[O II] / K & $n_e\,$[S II] / cm$^{-3}$& $12 + \log (\mathrm{O/H})$ & $\log (\mathrm{Ne/O})$ & $\log (\mathrm{Ar/O})$ \\
\midrule \midrule
EXCELS-40081 & $13600\pm1700$ & $13000\pm1800$ & $12500\pm1800$ & $254\pm161$ & $7.99\pm0.15$ & $-0.61\pm0.18$ & $-2.56\pm0.24$ \\
EXCELS-45052 & $11300\pm2400$ & $11100\pm2300$ & $10900\pm2100$ & $471\pm347$ & $8.38\pm0.30$ & $-0.57\pm0.36$ & $-2.77\pm0.45$ \\
EXCELS-45177 & $14100\pm1500$ & $13400\pm1700$ & $12900\pm1700$ & $659\pm255$ & $8.10\pm0.13$ & $-0.63\pm0.13$ & $-2.50\pm0.19$ \\
EXCELS-47143 & $13000\pm2300$ & $12500\pm2200$ & $12100\pm2100$ & $290\pm171$ & $8.18\pm0.23$ & $-0.62\pm0.21$ & $-2.64\pm0.27$ \\
EXCELS-52422 & $19700\pm2700$ & $18100\pm2500$ & $16800\pm2300$ & $300^a$ & $7.53\pm0.13$ & $-0.59\pm0.16$ & $-2.51\pm0.24$ \\
%EXCELS-52712 & $10500\pm1800$ & $10400\pm1800$ & $10400\pm1800$ & $876\pm352$ & $8.50\pm0.25$ & \textemdash$^{b}$ & $-2.86\pm0.29$ \\
EXCELS-59720 & $12500\pm1500$ & $12100\pm1700$ & $11700\pm1700$ & $970\pm376$ & $8.13\pm0.15$ & $-0.63\pm0.12$ & $-2.49\pm0.26$ \\
EXCELS-94335 & $11400\pm800$ & $11100\pm1300$ & $11000\pm1400$ & $339\pm97$ & $8.24\pm0.12$ & $-0.61\pm0.10$ & $-2.54\pm0.18$ \\
EXCELS-121806 & $14900\pm1300$ & $14100\pm1600$ & $13400\pm1600$ & $449\pm365$ & $7.95\pm0.10$ & $-0.55\pm0.09$ & $-2.30\pm0.17$ \\ \bottomrule
\end{tabularx}
\footnotesize $^{a}$ The density of EXCELS-52422 is not constrained as we do not detect the [S II] doublet, for this source we assume a fixed density of $300 \, \rm{cm}^{-3}$ (see Section~\ref{sec:analysis})
%\footnotesize $^{b}$ EXCELS-52712 lacks spectral coverage of \neiii \ and therefore no estimate for Ne/O.
\end{table*}

        To determine electron temperatures, densities and ionic abundances we use the \textsc{PyNeb} package \citep[v1.1.18;][]{Luridiana_2015}, adopting the atomic data listed in Table~\ref{tab:pyneb-data}.
        For each of our galaxies, we estimate the temperature of the high-ionization zone using the \oiiib /\oiiia \ ratio. 
        As there are no instances where we have a ${\rm S/N} > 2$ detection of the auroral \siiia \ emission line, we calculate the temperature of the intermediate ionization zone 
        using the $T_e \mathrm{[\textsc{O iii}]}$-$T_e \mathrm{[\textsc{S iii}]}$ scaling relation of \citet{Garnett_1992}, defined as  
        \begin{equation} \label{eq:garnett+1992}
           T_e[\textsc{S iii}] = 0.83 \times T_e\mathrm{[\textsc{O iii}]} + 1700 \ \mathrm{K},
        \end{equation}
        to which we add an uncertainty of $1300 \, \mathrm{K}$ in quadrature to account for the scatter of the relation \citep{Rogers_2021}. 
        Although we have a few significant detections of the \auroraloii \ line, for these sources we measure $T_e \mathrm{[\textsc{O ii}]} > > T_e \mathrm{[\textsc{O iii}]}$ which is unphysical. 
        The overestimated $T_e \mathrm{[\textsc{O ii}]}$ can be attributed to the high sensitivity of the  \auroraloii/\oii \ ratio to reddening corrections, flux calibration and density fluctuations \citep[e.g.,][]{Mendez_Delgado_2023}.
        Therefore, to ensure consistency for the full sample we use the $T_e \mathrm{[\textsc{O iii}]}$-$T_e \mathrm{[\textsc{O ii}]}$ scaling relationship of \citet{Campbell_1986}, defined as
        \begin{equation} \label{eq:campbell+1986}
            T_e[\textsc{O ii}] = 0.7 \times T_e\mathrm{[\textsc{O iii}]} + 3000 \ \mathrm{K},
        \end{equation}
        to measure the temperature of the low ionization zone, to which we add an additional uncertainty of $1100 \, \mathrm{K}$ \citep{Rogers_2021}. 

        \subsection{Electron density} \label{subsec:electron-density}
        
        Estimates of the electron densities ($n_e$) across the different ionization zones require measurements of emission-line doublets of the same ionic species, namely the \sii \ and \oii \ doublets for the low-ionization zone and the \arivdoublet \ doublet for the high-ionization zone \citep[e.g.,][]{Mingozzi_2022}.
        We use \siia/\siib \ as our density diagnostic as \oii \ is unresolved at the resolution of our data and we do not detect \arivdoublet \ in any of our galaxies.
        
        We assume a uniform density structure and use the $n_e\mathrm{[\textsc{S ii}]}$ estimate to calculate all ionic abundances.
        In general, this should not have a strong effect on our results.
        The $\mathrm{Ar^{2+}}$ and $\mathrm{O^{2+}}$ abundances are essentially insensitive to $n_e$ within the range $10^2 < n_e < 10^4 \, \mathrm{cm^{-3}}$, changing by $ \lesssim 0.001 \, {\rm dex}$ and $ \lesssim 0.005 \, {\rm dex}$ respectively.
        The $\mathrm{O^{+}}$ abundance is more sensitive to density, increasing significantly with density above $n_e \simeq 10^{3.5} \, {\rm cm^{-3}}$ (below this limit, the ionic abundance varies by $ \lesssim 0.02 \, {\rm dex}$, which is less than the uncertainty on a given oxygen abundance).
        However, this should not be a significant issue since our sample exhibits high \oiiinwl/\oiinwl ratios (i.e., log (\oiiia/\oii) $ = 0.5 \pm 0.2$) typical of high-redshift galaxies, indicating that $\mathrm{O^{2+}}$ dominates the total O abundance \citep[e.g.,][]{Shapley_2024}.
        Moreover, electron densities of $n_e > 10^{3.5} \, {\rm cm^{-3}}$ are rare for star-forming galaxies at the redshifts of our sample, and are instead typically within the range $100 - 500\, \mathrm{cm^{-3}}$ \citep[e.g.][]{Isobe_2023b}.
        
        In cases where the density is unconstrained (i.e., the \sii \ emission lines are not significantly detected), we assume a fixed $n_e = 300 \, \mathrm{cm^{-3}}$; this value is also broadly consistent with the inverse-variance weighted average value we determine from the EXCELS galaxies with $n_e$ constraints of $\simeq 300 \, \mathrm{cm^{-3}}$ (see Table~\ref{tab:abundances}).

        \subsection{Ionic and total abundances}

        We calculate the ionic abundances for each of our desired ionic species using the integrated line fluxes, the temperature corresponding to their ionization zone and the electron density.
        To account for uncertainties in the temperatures, densities and emission line strengths, we perturb each parameter by its associated error 500 times and re-calculate the abundances.
        The resulting ionic abundances and their uncertainties are given by the median and scaled median absolute deviation of the resulting distribution.
        
        Typically, the only observed ionization states of oxygen are $\mathrm{O^{+}}$ and $\mathrm{O^{2+}}$, with a negligible amount of oxygen in the $\mathrm{O^{3+}}$ state \citep[e.g.,][]{Berg_2021}.
        Therefore, the total oxygen abundance is simply sum the ionic abundances corresponding to $\mathrm{O^{+}}$ and $\mathrm{O^{2+}}$:
        \begin{equation} \label{eq:oxygen-abundance}
            \frac{\mathrm{O}}{\mathrm{H}} = \frac{\mathrm{O^{+}}}{\mathrm{H^{+}}} + \frac{\mathrm{O^{2+}}}{\mathrm{H^{+}}}.
        \end{equation}
        However, this is not the case for all elements of interest. 
        For example, we only observe the $\mathrm{Ar^{2+}}$ state of argon, and there are non-negligible contributions from $\mathrm{Ar^{+}}$ and $\mathrm{Ar^{3+}}$\footnote{Whilst we have spectral coverage of \arivdoublet, we have no significant detections of this feature in any of our galaxies. We note that including \arivnwl \ in the determination of Ar abundances typically results in differences of ${\log {\rm (Ar/O)}} \lesssim 0.05 \, {\rm dex}$ \citep{Arellano-Córdova_2024}, which is well within our uncertainties on our measured abundances.}.
        Similarly, we only observe the $\mathrm{Ne^{2+}}$ state of neon.
        To account for the unseen ions and infer the total abundance of these elements we employ an ionization correction factor (ICF) such that:
        \begin{equation} \label{eq:argon-abundance}
            \frac{\mathrm{X}}{\mathrm{H}} = \frac{\mathrm{X^{2+}}}{\mathrm{H^{+}}} \times \mathrm{ICF(X^{2+})}.
        \end{equation}
        From the variety of available ICFs, we use the ICFs of \citet{Izotov_2006} which are derived from photoionization modelling and incorporate dependencies on ionization parameter and metallicity.
        These ICFs have been shown to be suitable for high-redshift analogue galaxies \citep[e.g.,][]{Arellano-Córdova_2024} and, as such, should be appropriate for our high-redshift sample.
        We discuss the systematic uncertainties related to our choice of ICFs in Section~\ref{sec:icfs}.

\section{The Abundance Patterns of EXCELS galaxies}
\label{sec:results}

    In the following section, we present the ${\rm Ar/O}$ and ${\rm Ne/O}$ abundance ratios of our $1.8 < z < 5.3$ sample and interpret these results in the context of the contributions of CCSNe and SNe Ia.
    All relevant values of temperature, density, and element abundance are reported in Table~\ref{tab:abundances}.

    \subsection{Argon}

        \begin{figure}
            \centering
            \includegraphics[width=\linewidth]{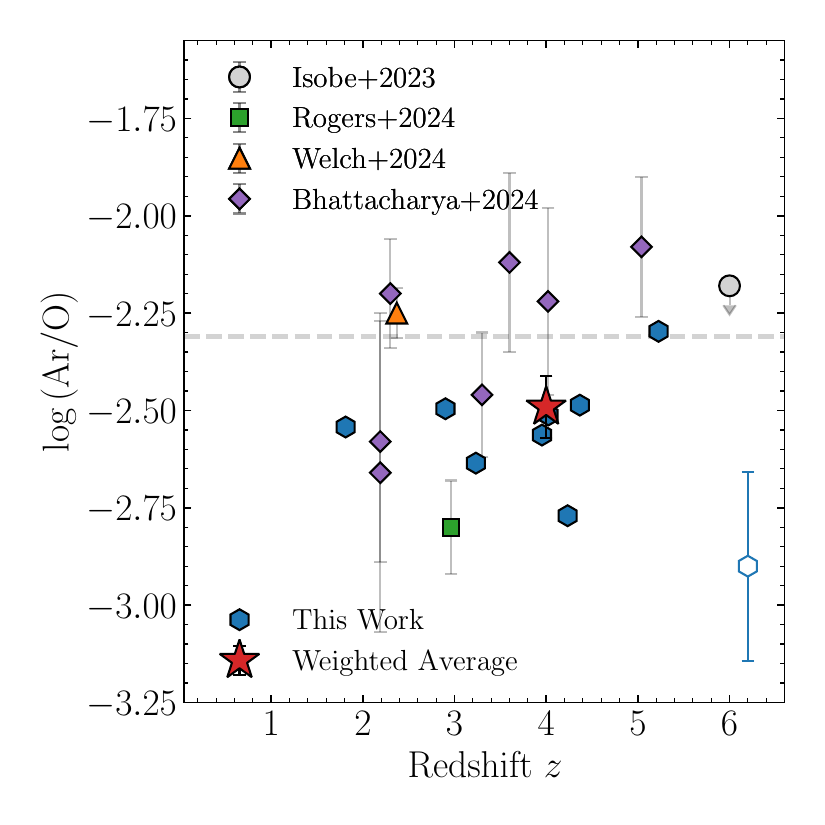}
            \caption{The $\log {\rm (Ar/O)}$ abundance as a function of redshift for our sample (blue hexagons) compared other to high-redshift samples from the literature. 
            We show the median average error bar in with an unfilled blue hexagon in the bottom right.
            The inverse-variance weighted average of our measurements is ${\log {\rm (Ar/O)} = -2.50 \pm 0.07}$ and is shown with a red star at the median redshift of the sample ($z=4.0$).
            For comparison, we show the measurements of \citet{Bhattacharya+2024} (purple diamonds), \citet{Rogers_2024} (green square) and \citet{Welch_2024} (orange triangle).
            We additionally plot an upper limit on $\log {\rm (Ar/O)}$ for a $z\simeq6$ galaxy with an electron temperature measurement from \citet{Isobe_2023}. 
            We denote the solar ratio, ${\log {\rm (Ar/O)_\odot}=-2.31}$, with a grey dashed line.
            The bulk of high-redshift measurements exhibit subsolar $\log {\rm (Ar/O)}$ ratios, indicating a deficiency in Ar enrichment in high-redshift star-forming galaxies.
            }
            \label{fig:ar-vs-redshift}
        \end{figure}  
        
        \begin{figure} 
            \centering
            \includegraphics[width=\linewidth]{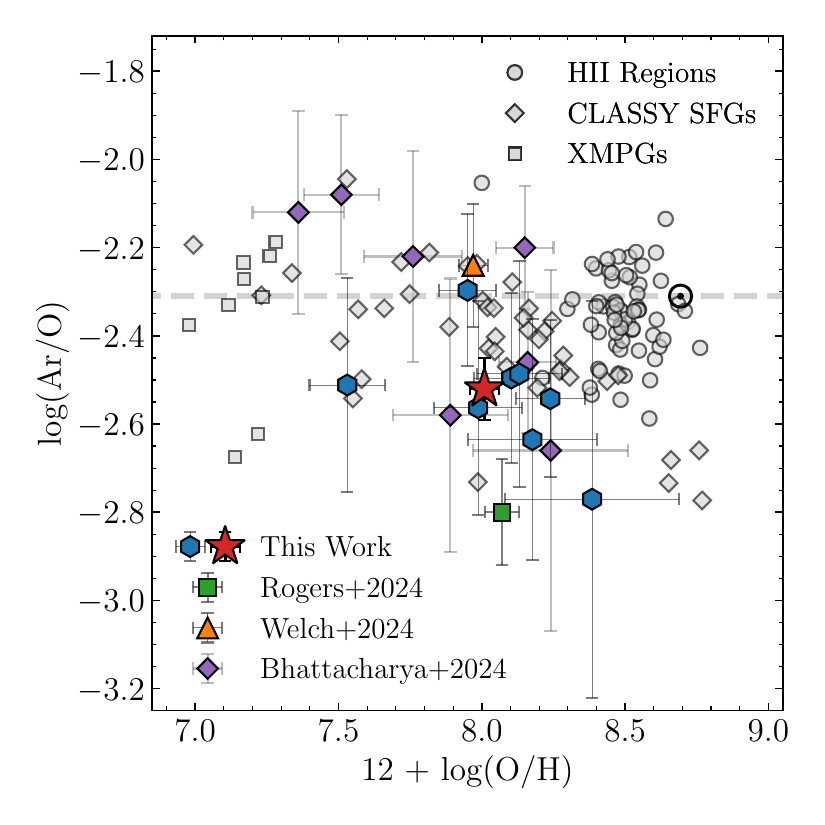}
            \caption{The ${\rm Ar/O}$ ratio as a function of ${\rm O/H}$ for the EXCELS galaxies (blue hexagons). 
            We compare to other high-redshift systems with reported ${\rm Ar/O}$ and ${\rm O/H}$ ratios using the same symbols as Fig.~\ref{fig:ar-vs-redshift}.
            In grey we show measurements from local \hii \ regions \citep[circles;][]{Rogers_2022}, extremely metal poor local galaxies from DESI \citep[squares; ][]{Zinchenko+2024}, and local analogues of high-redshift galaxies from CLASSY \citep[diamonds;][]{Arellano-Córdova_2024}.
            The solar value is denoted with the solar symbol and the solar ${\rm (Ar/O)_\odot}$ ratio with a grey dashed line. 
            The bulk of our sample lie below the solar $\log \mathrm{(Ar/O)_\odot}$ value and the local comparison sample.
            This suggests a deficiency of Ar with respect to O in high redshift systems.
            The inverse-variance weighted average of the EXCELS galaxies (red star) shows a $\simeq3.5\sigma$ offset from the solar value (such that $\rm{Ar/O} = 0.65 \pm 0.10 \, (\rm{Ar/O})_{\odot}$).
            These results indicate that the $\mathrm{Ar/O}$ ratio is sensitive to the star formation history.
            }
            \label{fig:aro-vs-oh}
        \end{figure}

\begin{figure*}
            \centering
            \includegraphics[width=0.8\linewidth]{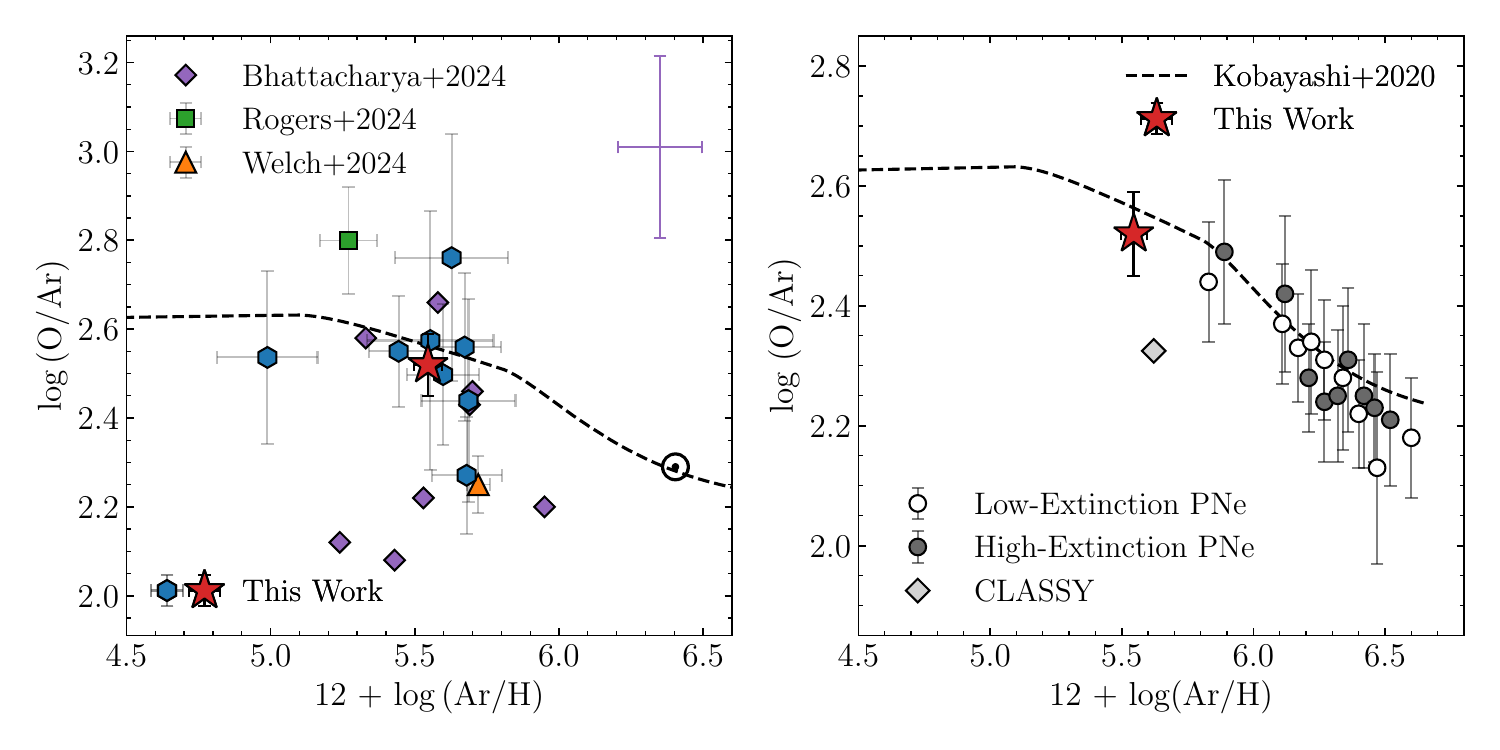}
            \caption{In the left panel we compare the ${\rm O/Ar}$ and ${\rm Ar/H}$ abundance ratios of our sample (blue hexagons) and the measurements of \citet{Bhattacharya+2024} (purple diamonds), \citet{Rogers_2024} (green square), and \citet{Welch_2024} (orange triangle) to the \citet{Kobayashi_2020} chemical evolution model representative of of the Milky Way.
            We show the median error of the \citet{Bhattacharya+2024} points in the top right.
            We denote the solar value using the solar symbol.
            To ensure an accurate comparison we have converted the \citet{Kobayashi_2020} out of solar units using their solar scale, namely $\log{\rm {(O/H)_\odot}} = 8.76$ and $\log{\rm {(Ar/H)_\odot}} = 6.40$. 
            The majority of the EXCELS galaxies are consistent with supersolar ${\rm O/Ar}$ ratios, including the inverse-variance weighted average of the sample (red star), indicating a deficit in Ar in these systems.
            The EXCELS galaxies generally agree with the trend shown by the \citet{Kobayashi_2020} relationship, with outliers attributed to more complex enrichment scenarios or poor performance of the ICF at high metallicity.
            In the right panel we show that the sample-weighted average is consistent with the trend set by the M31 PNe data (circles), the median abundance of the CLASSY galaxies (diamond), and the \citet{Kobayashi_2020} relationship, suggesting that high-redshift systems are on average evolving along similar pathways as systems such as the Milky Way.
            }
            
            \label{fig:aro-vs-arh}
        \end{figure*}

    In Fig.~\ref{fig:ar-vs-redshift} we show our measured ${\rm Ar/O}$ abundances as a function of redshift and compare to other literature results at $z>2$.
    All data points are derived from JWST/NIRSpec observations with a robust estimate of the electron temperature.
    It can be seen from our Fig.~\ref{fig:ar-vs-redshift} that our new sample significantly increases the number of galaxies with ${\rm Ar/O}$ estimates at these redshifts, adding eight galaxies to the thirteen reported measurements from the literature\footnote{We elect not to plot four of the galaxies from the \citet{Bhattacharya+2024} sample in Figs. \ref{fig:ar-vs-redshift}$-$\ref{fig:aro-vs-arh} based on potential contributions from active galactic nuclei or exotic types of supernovae.}.
    Fig. \ref{fig:ar-vs-redshift} also highlights the capability of moderately deep JWST spectroscopy ($\simeq 2-5$ hours integration time) to deliver Ar/O abundance estimates in the redshift range ${2 < z< 6}$.
    Future spectroscopic campaigns can significantly increase the number of such measurements and continue to build on the results presented here. 
    
    All but one galaxy in our EXCELS sample, and the majority of other measurements from the literature, fall below the solar ${\rm Ar/O}$ value \citep{Isobe_2023, Rogers_2024, Bhattacharya+2024}.
    A notable exception is the Sunburst Arc ($z=2.37$), which shows an approximately solar ratio with a relatively small statistical uncertainty \citep{Welch_2024}.
    However, it is clear that the average of all current estimates indicates an Ar deficit in high-redshift star-forming galaxies, which we discuss further below.
    It is perhaps interesting to note that all three current measurements at $z \gtrsim 5$ are consistent with solar or super-solar ${\rm Ar/O}$, which some authors have attributed to exotic enrichment sources at very early times \citep{Watanabe_2024, Bhattacharya+2024}.
    The one $z > 5$ galaxy in our sample (EXCELS-121806, $\log {\rm (M_\star \ M_\odot) = 8.11}$) has - perhaps by chance - the largest value of ${\rm Ar/O}$, and is consistent with the trend seen in other data.
    The sample size at these redshifts is obviously too small to draw any firm conclusions (one of the three estimates is also an upper limit) but more observations in this redshift regime are clearly of interest, especially given evidence for other anomalous abundance ratios at similar redshifts \citep[e.g. N/O;][]{Topping_2024}.
    
    \subsubsection{The metallicity dependence of Ar/O} \label{sec:ar_o-vs-o_h}

        In Fig.~\ref{fig:aro-vs-oh}, we show how the ${\rm Ar/O}$ ratio varies as a function of ${\rm O/H}$ (a proxy for total metallicity) in our sample.
        Alongside the high-redshift galaxy data, we also include a comparison to a selection of local data comprised of star-forming galaxies from the COS Legacy Archival Spectroscopic SurveY \citep[CLASSY; ][]{Berg_2022, James_2022, Arellano-Córdova_2024} at $0.02 < z < 0.18$ , extremely metal-poor galaxies from the early data release of the Dark Energy Spectroscopic Instrument (DESI) at $0.02 < z < 0.21$ \citep{Zinchenko+2024}, and extragalactic \hii \ regions in M33 from \citet{Rogers_2022}.
        
        The consensus of the local studies is that the ${\rm Ar/O}$ ratio remains approximately constant and consistent with the solar value at all ${\rm O/H}$, suggesting that the ${\rm Ar/O}$ ratio is insensitive to the history of star formation.
        Despite the metal-poor nature and low-stellar mass of some of these local sources, it is plausible that there has been enough time for SNe Ia enrichment which would serve to increase ${\rm Ar/O}$.
        For example, local dwarf galaxies and low-mass systems are known to exhibit less $\alpha$-enhancement (i.e., be more enriched in SNe Ia products) at fixed metallicity compared to stars in the Milky Way \citep[e.g. the SMC;][]{mucciarelli_2023}.
        
        In contrast to local samples, our EXCELS sample appears to be systematically offset to lower ${\rm Ar/O}$ at fixed ${\rm O/H}$.
        Although it is clear that the uncertainties on the individual measurements are large, the inverse-variance weighted average of the sample is ${\langle \log \mathrm{(Ar/O)} \rangle = -2.50 \pm 0.07}$ at ${\langle 12 + \log \mathrm{(O/H)} \rangle = 8.03 \pm 0.05}$, representing a $\simeq 3 \sigma$ deviation below the solar value.
        Our results are consistent with the $z\simeq3$ galaxy presented in \citet{Rogers_2024} which also shows a significant deviation to low ${\rm Ar/O}$.
        
        The other (albeit limited) high-redshift data show more mixed results, in particular the solar-like ${\rm Ar/O}$ derived from deep JWST/NIRSpec data of the Sunburst Arc \citep[$z=2.37$;][]{Welch_2024}.
        However, we note that this galaxy shares a number of properties similar to our highest ${\rm Ar/O}$ source EXCELS-121806 ($z = 5.226$; $\log ({\rm Ar/O}) = -2.30 \pm 0.17$) including its total metallicity and evidence of enhanced N abundance (discussed in a companion paper; Arellano-Córdova et al. in prep).
        This could suggest that high ${\rm Ar/O}$ values in fact represent outliers in the high-redshift population related to enrichment scenarios that are not yet fully understood, but much larger samples are needed before this can be robustly addressed.
        The larger ${\rm Ar/O}$ derived by \citet{Welch_2024} may also be related to systematic uncertainties in the temperature structure of high-redshift \hii \ regions, which we discuss further in Section \ref{subsec:temperature-structure-uncertainty}.
        
        The \citet{Bhattacharya+2024} sample shows no strong evidence for systematically lower ${\rm Ar/O}$, however, we note that their estimates are derived from much shallower data compared to all other high-redshift samples.\footnote{The majority of the \citet{Bhattacharya+2024} spectra have exposure times of $< 3$ hours per grating, with a minimum exposure time of $0.85$ hours for the CEERS targets. In comparison, the EXCELS spectra have $4$ hours in the G140M/G395M gratings and $5.5$ hours in the G235M grating.}
        As such, \citet{Isobe_2023} have reported upper limits for a number of these sources.
        Overall, our observations presented here suggest that the ${\rm Ar/O}$ ratios of high-redshift galaxies are most likely sub-solar, consistent with expectations if Ar has significant SNe Ia production pathway.
        This is in contrast to local systems, which generally show higher ${\rm Ar/O}$.
        In the following section, we will discuss these results in more detail via a comparison to a state-of-the-art chemical evolution model.
        
        Finally, we note that one galaxy in our sample, EXCELS-45052, appears highly deficient in Ar with respect to O, with an Ar/O ratio of $\simeq -2.8$.
        This galaxy is metal-enriched, with $12 + \log {\rm (O/H) = 8.38 \pm 0.30}$, and thus falls within the `high-metallicity' regime ($12 + \log \mathrm{(O/H)} \geq 8.2$) within which the Ar ICF has been observed to have a strong metallicity dependence and large dispersion of $\simeq 0.16$ dex \citep{Arellano-Córdova_2024}.
        Therefore, it is possible this significantly sub-solar ${\rm Ar/O}$ ratio is an artefact of an unreliable ICF at high metallicity.
        Further investigation of the Ar ICFs are required to fully address this issue.
        We discuss these systematics more thoroughly in Section~\ref{sec:dust-depletion}.
        However, we note that, given its significant uncertainty, this galaxy does not strongly bias the weighted average of our sample.

        \begin{figure}
            \centering
            \includegraphics[width=\linewidth]{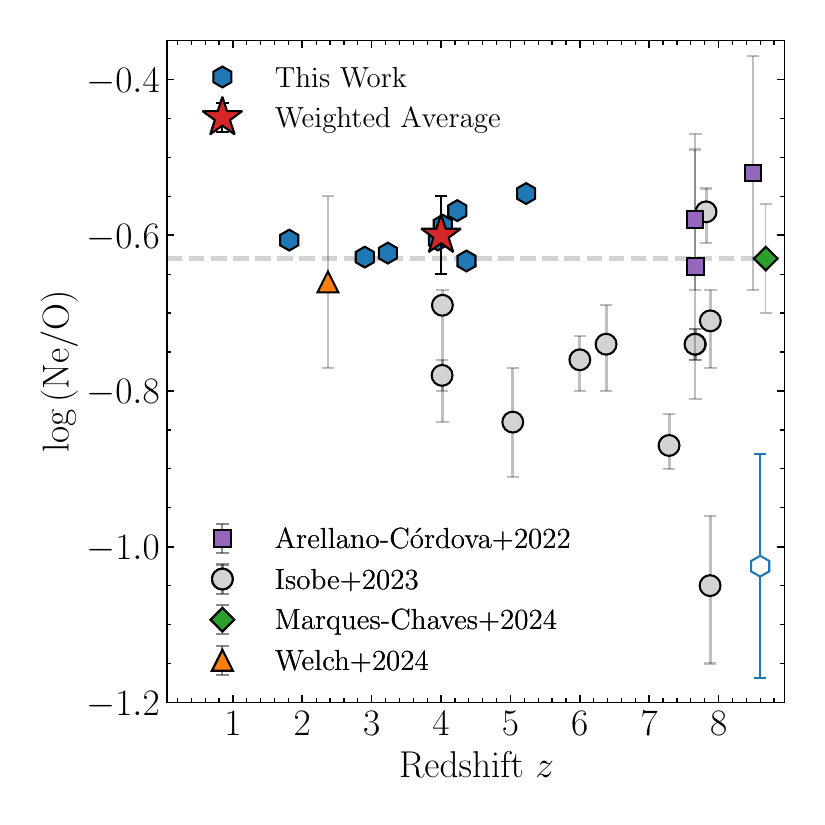} 
            \caption{The $\log {\rm (Ne/O)}$ abundance as a function of redshift for our sample (blue hexagons).
            We show the median average error bar in with an unfilled blue hexagon in the bottom right.
            For comparison, we show the direct-method measurements of \citet{Arellano-Cordova_2022} (purple squares), \citet{Isobe_2023} (grey circles), \citet{Marques-Chavez_2024} (green diamond) and \citet{Welch_2024} (orange triangle).
            The inverse-variance weighted average of our measurements, ${\log {\rm (Ne/O)} = -0.60 \pm 0.05}$, is shown with a red star at the median redshift of the sample.
            We denote the solar ratio, ${\log {\rm (Ne/O)_\odot}=-0.63}$, with a grey dashed line.
            Our Ne/O ratios show excellent consistency with the solar relation and local samples, suggesting O and Ne are enriched via similar pathways.
            }
            \label{fig:ne-vs-redshift}
        \end{figure}

        \subsubsection{A comparison to chemical evolution models and inferences from M31 planetary nebulae} \label{sec:o_ar-vs-ar_h}

        In this section, and in Fig.~\ref{fig:aro-vs-arh}, we compare our results to the Milky Way chemical evolution model of \citet{Kobayashi_2020}.
        We adopt the convention of plotting the ratio of the pure CCSNe element to the SNe Ia element, i.e., $\log {\rm (O/Ar)}$, as a function of the SNe Ia element.
        This framing is directly analogous to the $\log {\rm (O/Fe)} - \log {\rm (Fe/H)}$ diagram which have previously been employed in studies of $\alpha$-enhancement in high-redshift galaxies \citep[e.g.][]{Cullen_2021, Kashino_2022, Chartab_2023, Stanton_2024}.
        
        The \citet{Kobayashi_2020} model (dashed line in Fig.~\ref{fig:aro-vs-arh}) can be interpreted as a temporal relation, with the plateau at ${\rm 12 + \log(Ar/H)} \lesssim 5.0$ corresponding to CCSNe-dominated yields at early times, the 'knee' at ${\rm \log (Ar/H)} \simeq 5.5$ corresponding to the onset of SNe Ia, and the decline at ${\rm \log (Ar/H)} \gtrsim 6.0$  corresponding to the steady build up of SNe Ia products at late times.
        It can be seen that the EXCELS sample, and most other high-redshift sources, are in good agreement with the \citet{Kobayashi_2020} model, generally falling close to the knee of the relation.
        Interestingly, this situation is consistent with the locations of $z\simeq3.5$ star-forming galaxies in the $\log {\rm (O/Fe)} - \log {\rm (Fe/H)}$ plane, which are also observed to lie close to, or just beyond, the knee corresponding to the onset of SNe Ia enrichment \citep[e.g.][]{Cullen_2021, Stanton_2024}.
        This aligns with the expectation that young star-forming galaxies are yet to undergo significant SNe Ia enrichment.
        However, it is important to note that the \citet{Kobayashi_2020} model is not universal but is built specifically to match the abundance patterns observed in Milky Way stars.
        Galaxies falling above/below the \citet{Kobayashi_2020} model (e.g., EXCELS-121806 or the measurement of \citealp{Welch_2024}) could be explained by assuming a different star-formation history.
        As an example, an infall of pristine gas between starbursts could serve to maintain the ${\rm \log (O/Ar)}$ abundance ratio but decrease the ${\rm log (Ar/H)}$ abundance, leading to galaxies residing below the Milky Way relation \citep{Bhattacharya+2024}
        Nevertheless, this comparison demonstrates consistency between our data and the \emph{expectations} for low-metallicity galaxies at early epochs under a sensible star-formation history assumption.

        Furthermore, we find that our measurements are consistent with observations in the local Universe.
        In the right panel of Fig.~\ref{fig:aro-vs-arh}, we compare the weighted-average abundances of our sample to measurements of planetary nebulae (PNe) in M31 from \citet{Arnaboldi+2022}.
        The PNe closely follow the decreasing branch of the \citet{Kobayashi_2020} model, indicative of an increase in SNe Ia enrichment causing a decrease in ${\rm O/Ar}$ at higher metallicity (i.e., increasing ${\rm Ar/H}$).
        Our sample average lies at the natural extension of this sequence, falling closer to the knee of the relationship, reflecting an epoch of relatively little SNe Ia enrichment.
        The inverse-variance weighted abundances of the CLASSY sample fall below the MW chemical evolution model relationship.
        However, it is plausible that the star-formation efficiencies of the dwarf galaxies in CLASSY are lower than is typical for the MW and the galaxies in our sample; at lower star-formation efficiencies the knee in the chemical evolution mode would occur at lower Ar/H, leading to lower O/Ar at fixed Ar/H.
        Indeed, in a companion paper we find evidence for differences in the typical star-formation histories of high-redshift galaxies and CLASSY galaxies based on their CNO abundance patterns \citep{arellano-cordova-2024}.
        
        These comparisons may suggest that, on average, high-redshift systems follow the same evolutionary pathways as traced by the Milky Way and M31, although direct comparisons such as these are not straightforward (Monty et al. in prep).
        At the very least, however, the consistent picture emerging between local observations, high-redshift observations, and the predictions of chemical evolution models, as suggested by Fig.~\ref{fig:aro-vs-arh}, is reassuring.
                
        \subsection{Neon} \label{sec:ne/o} 
        
        \begin{figure}
            \centering
            \includegraphics[width=\linewidth]{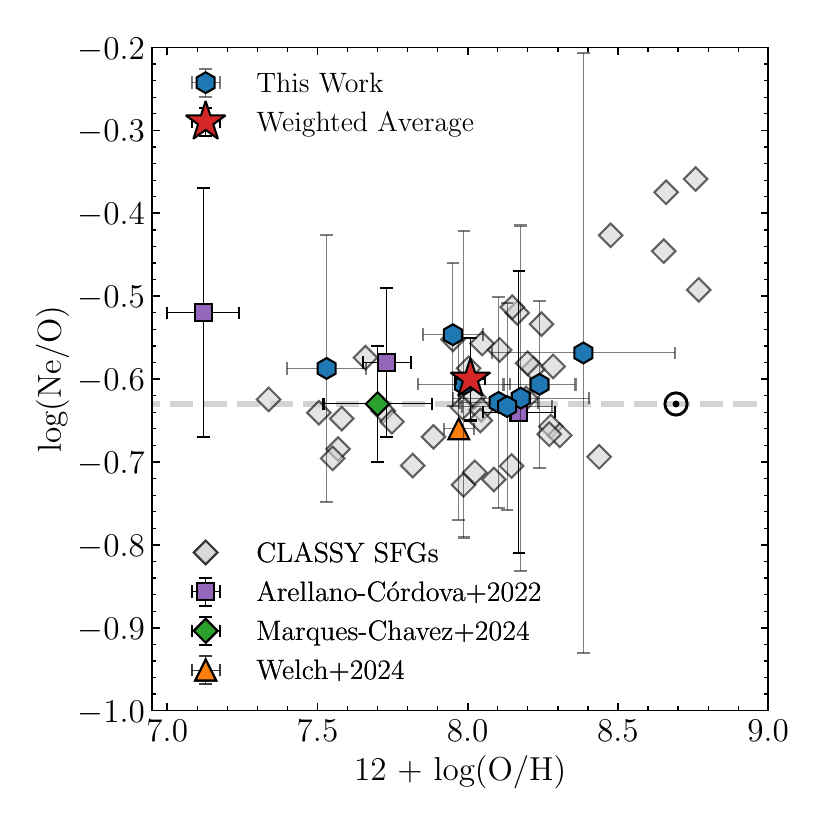}
            \caption{
            The ${\rm Ne/O}$ abundances of the EXCELS galaxies (blue hexagons) as a function of metallicity.
            For comparison, we plot the high-redshift measurements of \citet{Arellano-Cordova_2022} (purple squares), \citet{Marques-Chavez_2024} (green diamond) and \citet{Welch_2024} (orange triangle).
            The local analogue CLASSY sample \citep{Arellano-Córdova_2024} is plotted in grey diamonds, and we denote the solar abundance with $\odot$, and the ${\rm (Ne/O)_\odot}$ ratio with a grey dashed line.
            The inverse-variance weighted average of of our sample (red star) is $\langle \log {\rm (Ne/O)} = -0.60 \pm 0.05$ and is consistent with the solar value within $1\sigma$.
            Across the entire ${\rm \log (O/H)}$ range, the ${\rm Ne/O}$ ratio appears broadly constant and highly consistent with the solar value and local samples.
            }
            \label{fig:neon-oxygen}
        \end{figure}  
        
        Ne is an example of a typical $\alpha$-element which is expected to be produced primarily by CCSNe with no production channel through SNe Ia \citep{Kobayashi_2020}.
        Although there is some evidence for excess Ne enrichment at high metallicities ($12 + \log {\rm (O/H)} \gtrsim 8.5$; \citealp{Kobayashi_2020, Miranda_Perez_2023, Arellano-Córdova_2024}), this should not be relevant for our low-metallicity $z\simeq4$ sample.
        As such, Ne acts as a useful control element; in contrast to ${\rm Ar/O}$, the ${\rm Ne/O}$ ratio is \emph{not} expected to evolve across cosmic time \citep[e.g., ][]{Arellano-Cordova_2022}.
        
        In Fig.~\ref{fig:ne-vs-redshift} and Fig.~\ref{fig:neon-oxygen} we show ${\rm Ne/O}$ as a function of redshift and $12 + \log {\rm (O/H)}$ (i.e., mirroring Figs.~\ref{fig:ar-vs-redshift} and \ref{fig:aro-vs-oh} for Ar).
        From Fig.~\ref{fig:ne-vs-redshift} it can be seen that our sample and the majority of other high-redshift estimates are consistent with the solar value \citep[e.g.,][]{Arellano-Cordova_2022, Marques-Chavez_2024, Welch_2024}.
        The inverse-variance weighted average of our new EXCELS sample is $\langle \log \rm{(Ne/O)} \rangle = -0.60 \pm 0.05$, which is fully consistent with the solar value within $ < 1 \sigma$.
        
        These results are also fully consistent with the results of \cite{Arellano-Córdova_2024}, who find an unweighted average of $\langle \log \rm{(Ne/O)} \rangle = -0.63 \pm 0.06$ for local galaxies in the CLASSY survey.
        It is worth noting that Ne/O measurements from local star-forming galaxies typically fall systematically below the solar value.
        For example, \citet{Izotov_2006} find $\langle \log \rm{(Ne/O)} \rangle \simeq -0.75$ for a sample of metal-poor galaxies from SDSS.
        However, \citet{Arellano-Córdova_2024} re-analysed these ${\rm Ne/O}$ ratios with updated atomic data \citep[see also][]{Berg_2015}, finding results consistent with the solar value ($\langle \log \rm{(Ne/O)} \rangle \simeq -0.65$). 
        
        However, \citet{Isobe_2023} report generally subsolar ${\rm Ne/O}$ abundances for their galaxies at $4 < z < 8$.
        Their lowest estimates of $\log {\rm (Ne/O)} < -1.0$ can be attributed to the presence of massive ($> 30 {\rm M_\odot}$) stellar populations \citep[see][]{Watanabe_2024}.
        However, it is unclear what drives their other ${\rm Ne/O}$ abundances to lower values than measured from other galaxy samples at similar redshifts.
        In either case, the majority of Ne/O values at high redshift are consistent with local estimates.
        Fig.~\ref{fig:neon-oxygen} further highlights this point, where it can be seen that the $\rm{Ne/O}$ values we derive align with local star-forming galaxies from \citet{Arellano-Córdova_2024} at all metallicities, in contrast to the offset in $\rm{Ar/O}$ shown in Fig. \ref{fig:aro-vs-oh}.

\begin{figure} 
            \centering
            \includegraphics[width=0.7\linewidth]{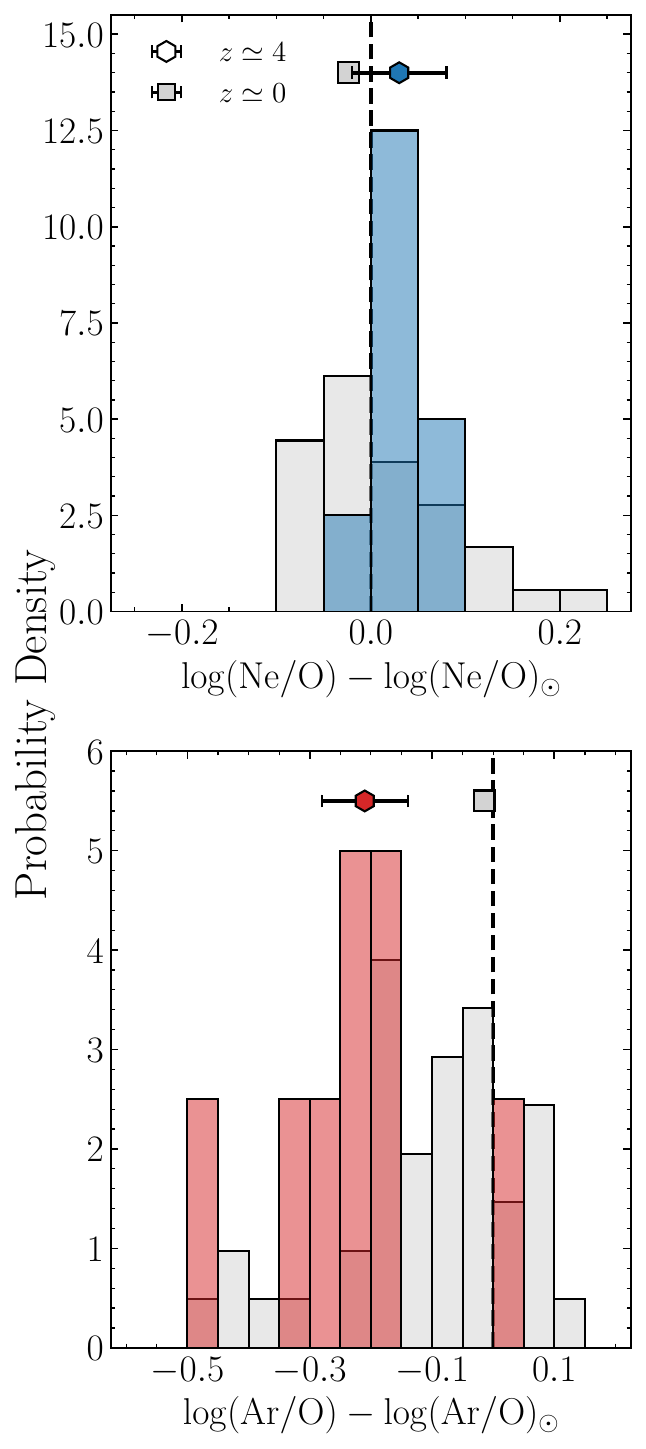}
            \caption{Probability density histograms highlighting the differences in our abundances from local samples.
            In each panel we subtract the corresponding solar value to more clearly show the difference in each abundance ratio.
            In the top panel we compare the ${\rm Ne/O}$ measurements of our sample (blue) to the $z\simeq0$ sample of CLASSY \citep[grey;][]{Arellano-Córdova_2024}.
            The inverse-variance weighted averages of each are shown by a blue hexagon and grey square respectively.
            In the bottom panel we show the same bur for ${\rm Ar/O}$, with our sample coloured red in this case.
            The ${\rm Ne/O}$ ratio clearly shows no clear evolution at $z\simeq4$ compared to local star-forming galaxies, with fully consistent averages.
            In contrast, ${\rm Ar/O}$ shows an offset to lower values at higher redshifts.
            These trends are consistent with a scenario wherein Ne is enriched through the same channels as O, whilst Ar has a delayed secondary enrichment through SNe Ia. 
            }
            \label{fig:diff-from-solar}
        \end{figure}

        Our observations therefore support a scenario in which Ne acts as a pure $\alpha$ element enriched mainly by CCSNe, whereas Ar has an additional - and significant - production channel via SNe Ia.
        As a result, local and high-redshift star-forming galaxies display similar $\rm{Ne/O}$  ratios, whereas the $\rm{Ar/O}$ ratio is lower in high-redshift systems as a result of the delayed onset of SNe Ia.
        This basic result is illustrated in Fig.~\ref{fig:diff-from-solar}, where we show the distributions of $[\rm{Ne/O}]$ and $[\rm{Ar/O}]$ for our sample compared to the CLASSY sample at $z \simeq 0$ \citep{Arellano-Córdova_2024}.
        These results are in line with expectations based on the expected star-formation histories of $z\simeq4$ star-forming galaxies combined with CCSNe and SNe Ia yield models \citep[e.g.][]{Kobayashi_2020}.
        However, it is clear that the current sample sizes are relatively small.
        Moreover, as discussed in Section \ref{sec:data}, the sample presented here is biased to above average SFRs compared to the full $z\simeq4$ star-forming population.
        Therefore, more complete samples with direct O, Ne, and Ar abundance estimates are required to confirm that these trends are characteristic of the full galaxy population at $2 < z < 6$.

\section{Discussion}
\label{sec:discussion}

    In the following section, we discuss some important systematic uncertainties that may affect our measured abundance ratios, namely the assumed temperature structure, dust depletion, and the choice of ICF and atomic data.
    We limit our discussion here to the systematics as they relate to Ar abundances (although the same considerations also apply to Ne and, to a lesser extent, O).

    \subsection{Atomic data and ICFs} \label{sec:icfs}

        Our chosen set of atomic data (see Table~\ref{tab:pyneb-data}) follows similar studies of high-redshift star-forming galaxies \citep[e.g.,][]{Rogers_2024}.
        However, alternative data on transition probabilities and collision strengths exist.
        In particular, our main $z\simeq0$ reference sample (CLASSY; \citealp{Arellano-Córdova_2024}) use the collision strengths of \citet{Galavis_1995} and the transition probabilities of \citet{Mendoza_1983} and \citet{Kaufman_1986} for ${\rm Ar^{2+}}$ in their analysis.
        Switching to these atomic data, the average difference ($\pm$ standard deviation) of the $\log {\rm (Ar/H)}$ abundances we derive is $0.04 \pm 0.09 \, {\rm dex}$.
        This change is insufficient to align the average Ar abundance of our sample with the $z\simeq0$ data and solar value.

        We also explore different choices for the Ar ICF, considering the ICFs of \citet{Pérez-Montero_2007} and \citet{Amayo_2021}.
        Applying either of these ICFs results in a slight increase in our measured Ar abundances; the \citet{Pérez-Montero_2007} ICF increases $\log {\rm (Ar/H)}$ by $0.04 \pm 0.09 \, {\rm dex}$ whilst the \citet{Amayo_2021} ICF results in an increase of $0.04 \pm 0.13 \, {\rm dex}$.
        Both of these offsets are again minor, and the resulting abundances are fully consistent with the abundances using the \citet{Izotov_2006} ICF, and thus the deviation of ${\rm Ar/O}$ from the solar value cannot be attributed to the choice of ICF.
        Taking combinations of the alternative ICFs and atomic data yields $\langle \log {\rm (Ar/O)} \rangle < -2.40$ which is still $\simeq 2\sigma $ discrepant from the solar value.
            
    \subsection{Dust Depletion} \label{sec:dust-depletion} 

        Dust in the ISM can deplete heavy elements from the surrounding gas, with the degree of depletion depending on the condensation temperatures of each element \citep{Savage_Sembach_1996, Groves_2004}.
        Both Ar and Ne are noble gases with low condensation temperatures (i.e., $47 \, {\rm K}$ and $9 \, {\rm K}$) and, as such, should not condense and become trapped in dust grains within \hii \ regions \citep{Lodders_2003, Sofia_2004}.
        However, O has a higher condensation temperature ($180 \, {\rm K}$) and has been observed to deplete onto dust.
        For example, \citet{Peimbert+Peimbert_2010} measured O abundances in galactic and extragalactic \hii \ regions, finding that O-depletion onto dust grains in systems with $7.3 < 12 + \log \mathrm{(O/H)} < 7.8$ decreases the measured $ \rm{log} \mathrm{(O/H)}$ by an average of $0.09 \pm 0.03 \, \mathrm{dex}$ ($\simeq 25$ per cent).
        Higher metallicity systems in the range $8.3 < 12 + \log \mathrm{(O/H)} < 8.8$ have a slightly larger fraction of their oxygen trapped in dust ($0.11 \pm 0.03 \, \mathrm{dex}$).
        Further analysis by \citet{Peña-Guerrero_2012} finds that estimation of the true oxygen abundance of a system requires increasing the oxygen abundance by $\simeq 0.10 \, \mathrm{dex}$.
        Similarly, \citet{Gutkin_2016} found that in order to accurately reproduce the expected optical emission line diagrams of star-forming galaxies in their models required a $\simeq 30$ per cent depletion of O onto dust grains.

        Throughout this work, we have followed the methodology of our main comparison samples and have chosen not to correct for the effect of O depletion.
        However, applying a dust correction would decrease the ${\rm Ar/O}$ ratio by $0.10 \, \mathrm{dex}$.
        This correction would only serve to increase the separation between the solar ${\rm Ar/O}$ ratio and local galaxies and thus strengthen our conclusions. 
        A similar correction to the ${\rm Ne/O}$ ratio would bring our average below the solar ratio, but still consistent within $< 2.3 \sigma$.
        Therefore, our results would not change - and if anything would be strengthened - by including dust depletion.
        
    \subsection{Temperature structure} \label{subsec:temperature-structure-uncertainty}

        The direct method approach for measuring chemical abundances relies on accurate temperature and density estimates.
        As discussed in Section~\ref{subsec:electron-density}, our abundance measurements are relatively insensitive to the exact density value within the low-density regime of our sample (i.e., as long as the density does not exceed $\mathrm{log}(n_e/\mathrm{cm}^{-3}) \simeq 3.5$).
        However, systematic differences in the electron temperatures assigned to each ionization zone can have a significant effect on the measured abundances.
        The fundamental, and mostly unavoidable, problem for high-redshift observations is that we often only have access to the high-ionization zone temperature via the auroral \oiiib \ line.
        In this situation, it is common practice, as we have done, to apply temperature scaling relationships to infer the temperatures of the other ionization zones.
        These relationships are derived either from photoionization models \citep[e.g.,][]{Campbell_1986, Garnett_1992} or low-redshift observations \hii \ regions \citep[e.g.,][]{Croxall_2016, ArellanoCordova_Rodriguez_2020, Rogers_2021}.
        However, whether these temperature scaling relations accurately describe \hii \ regions in high-redshift galaxies has still not been robustly verified.

        The recent analyses of \citet{Welch_2024} and \citet{Rogers_2024} serve to illustrate this issue. 
        In both cases, access to deeper data allows these authors to measure the intermediate ionization zone temperature using the \siiia \ auroral line.
        \citet{Rogers_2024} find a value consistent with local empirical relations and photoionization models, in which $T_e\rm{[SIII]} \simeq T_e\rm{[OIII]}$.
        In contrast, \citet{Welch_2024} find $T_e\mathrm{[SIII]} << T_e\mathrm{[OIII]}$ for the Sunburst Arc.
        Indeed, this is the main reason why \citet{Welch_2024} derived a solar-like ${\rm Ar/O}$ ratio.
        Applying $T_e\rm{[SIII]} \simeq T_e\rm{[OIII]}$ to the \citet{Welch_2024} data yields ${\log {\rm (Ar/O)}} \simeq -2.4$, more consistent with our analysis.

        It is clear that the assumed temperature structure can cause significant variation in the measured abundances.
        Future JWST spectroscopic observations, with sufficient depth to robustly measure the temperatures of multiple ionization zones, are needed to test the validity of the commonly used $T_e-T_e$ relations at high redshift.
        A caveat to the results presented here is that we \emph{assume} current $T_e-T_e$ relations are valid at $z\simeq4$.

\section{Conclusions}
\label{sec:conclusions}

    In this work we have presented direct O, Ar and Ne abundance measurements for eight star-forming galaxies in the redshift range $1.8 < z < 5.3$ from the JWST EXCELS survey \citep{Carnall_2024}.
    We have compared our sample with other high redshift and local samples from the literature to investigate the relationships between ${\rm Ar/O}$, ${\rm Ne/O}$ and ${\rm O/H}$.
    The primary motivation of the study has been to determine whether there is evidence for an Ar deficit in high-redshift galaxies as a result of significant Ar production via SNe Ia. 
    Our main results are as follows:

    \begin{enumerate}
    
        \item We find that the bulk of our sample exhibit Ar/O values below the solar value of ${\log \mathrm{(Ar/O}_{\odot}) = -2.31}$ (Figs.~\ref{fig:ar-vs-redshift} and \ref{fig:aro-vs-oh}). 
        The weighted average of our measurements is ${\langle  {\rm \log (Ar/O)} \rangle = -2.50 \pm 0.07}$, a ${\simeq 3\sigma}$ offset below solar.
        On a linear scale, our result is $\langle {\rm{Ar/O} \rangle = 0.65 \pm 0.10 \, (\rm{Ar/O})_{\odot}}$, which is consistent with supernova yield models that predict $\simeq 34$ per cent of Ar production occurs via SNe Ia, and implies that the ISM of star-forming galaxies at $z\simeq4$ are dominated by CCSNe products.

        \item In contrast, we find no deviation from the solar value when considering the ${\rm Ne/O}$ ratio (Figs.~\ref{fig:ne-vs-redshift} and \ref{fig:neon-oxygen}).
        We find a weighted average Ne abundance of ${\langle \rm{Ne/O} \rangle = 1.07 \pm 0.12 \, (\rm{Ne/O})_{\odot}}$.
        This result implies that Ne can be considered a pure $\alpha$-element that follows the same enrichment pathways as O, as expected by models.
        
        \item We have verified our results against a sample of local star-forming galaxies with similar total metallicity from the CLASSY survey (\citealp{Arellano-Córdova_2024}; Figs.~\ref{fig:aro-vs-oh}, \ref{fig:neon-oxygen} and \ref{fig:diff-from-solar}).
        Our sample shows similar Ne/O but lower Ar/O compared to these local star-forming analogues.
        
        \item We have compared our measurements to the Milky Way chemical evolution model of \citet{Kobayashi_2020} as well as local measurements of PNe in M31 (Fig.~\ref{fig:aro-vs-arh}).
        Our sample is generally in good agreement with the model predictions, which can be again interpreted as a lack of SNe Ia enrichment in high-redshift star-forming galaxies.
        We find that our average Ar/O estimate also lies at the extension of the sequence inferred from PNe in the [O/Ar]-[Ar/H] plane (Fig. \ref{fig:aro-vs-arh}).
        Taken together, these comparisons imply that the Ar deficit we observe is consistent with \emph{both} local data and the predictions chemical evolution models.
                
        \item We have assessed the effect of systematic uncertainties related to ICFs, atomic data, and dust depletion on our main result.
        We find that, when combined, these affects can alter ${\rm \log (Ar/O)}$ at the $\simeq 0.1$ dex level, however there is no strong evidence to suggest that this would alter our main conclusions.
        Uncertainties in the temperature structure of high-redshift \hii \ regions can potentially have a significant effect (as illustrated by the \citealp{Welch_2024} result) and verifying $T_e-T_e$ relations should be a high priority of future JWST programmes.
        
    \end{enumerate}

    This work constitutes a significant increase in the sample size of galaxies with robust Ar abundances beyond $z > 2$, and our initial results are consistent with a picture in which Ar is deficient in high-redshift star-forming galaxies due to delayed SNe Ia enrichment.
    These results corroborate independent analyses of the O/Fe ratio at similar redshifts \citep{Steidel_2016, Cullen_2019, Topping_2020A, Kashino_2022, Chartab_2023, Stanton_2024}.
    However, the current number of galaxies with direct Ar abundances measurements remains small, and additional data, as well as continued work on ICFs, dust depletion and the temperature structure of high-redshift \hii \ regions are still required to robustly confirm these trends.
    Nevertheless, our analysis demonstrates the potential for using JWST to unravel the detailed chemical enrichment patterns observed in the early Universe.

\section*{Acknowledgements}

    T. M. Stanton, F. Cullen, K. Z. Arellano-C\'ordova and D. Scholte acknowledge support from a UKRI Frontier Research Guarantee Grant (PI Cullen; grant reference: EP/X021025/1). 
    A. C. Carnall and S. Stevenson acknowledges support from a UKRI Frontier Research Guarantee Grant (PI Carnall, grant reference EP/Y037065/1). 
    R. Begley, C. Bondestam, C. T. Donnan, J. S. Dunlop, D. J. McLeod  and R. J. McLure acknowledge the support of the Science and Technology Facilities Council. 
    J. S. Dunlop also thanks the Royal Society for their support through a Royal Society Research Professorship. 
    This research made use of the following software: \textsc{matplotlib} \citep{Hunter_2007}, \textsc{pyneb} \citep{Luridiana_2015}, \textsc{numpy} \citep{Harris_2020}, \textsc{dynesty} \citep{Speagle_2020} and \textsc{msaexp} \citep{Brammer_2023}.
    This research utilised NASA's Astrophysics Data System Bibliographic Services.
    For the purpose of open access, the author has applied a Creative Commons Attribution (CC BY) licence to any Author Accepted Manuscript version arising from this submission. % for later

%%%%%%%%%%%%%%%%%%%%%%%%%%%%%%%%%%%%%%%%%%%%%%%%%%
\section*{Data Availability}

    All data will be shared by the corresponding author upon reasonable request.
    

\bibliographystyle{mnras}
\bibliography{excels-ar-o.bib} 

%%%%%%%%%%%%%%%%% APPENDICES %%%%%%%%%%%%%%%%%%%%%

\appendix

\section{Zoom-ins on \texorpdfstring{\oiiib \ and \ariii}{auroral oxygen and argon}}

        \begin{figure*}
            \centering
            \includegraphics[width=0.9\textwidth]{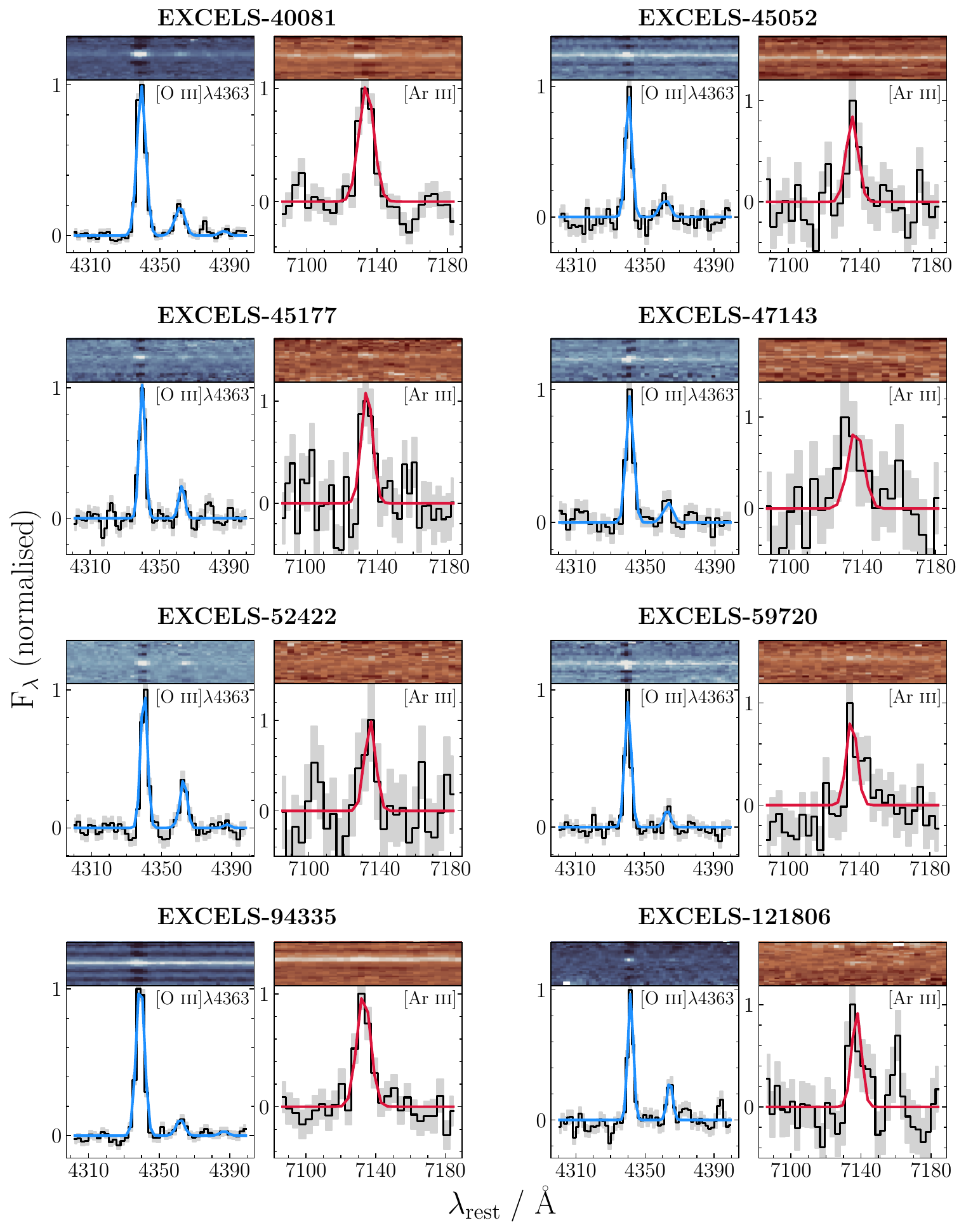}
            \caption{
            Zoom-ins on the principal faint emission lines of our sample.
            The continuum-normalised underlying spectrum is plotted in black, with the uncertainty on the spectrum is shaded in grey.
            We normalise each window by the maximal flux within its corresponding spectral region.
            \hgamma \ and \oiiib \ fits are shown in blue and \ariii \ fits in red.
            Above each window, we show the corresponding 2D spectra.
            }
            \label{fig:emission-line-fits}
        \end{figure*}
        
        In Figure~\ref{fig:emission-line-fits}, we show zoom ins on the individual emission line fits for the key emission lines for this study.
        Each window is continuum normalised with the best fitting profile overplotted.
        Above each line, we show a cut out of the 2D spectrum.

%%%%%%%%%%%%%%%%%%%%%%%%%%%%%%%%%%%%%%%%%%%%%%%%%%

% Don't change these lines
\bsp	% typesetting comment
\label{lastpage}
\end{document}